\documentclass[aps,twocolumn,superscriptaddress,floatfix]{revtex4}
\usepackage{epsfig}
\usepackage{graphicx}
\usepackage{amssymb}

\begin{document}
\title{Test of classical nucleation theory on deeply supercooled
  high-pressure simulated silica}

\author{Ivan Saika-Voivod}
\affiliation{Department of Chemistry, University of Saskatchewan,
Saskatoon, Saskatchewan, S7N 5C9, Canada}

\author{Peter H. Poole}
\affiliation{
Department of Physics, St. Francis Xavier University, Antigonish, NS,
B2G 2W5, Canada}

\author{Richard K. Bowles}
\affiliation{Department of Chemistry, University of Saskatchewan,
Saskatoon, Saskatchewan, S7N 5C9, Canada}

\date{\today}

\begin{abstract}
  We test classical nucleation theory (CNT) in the case of simulations
  of deeply supercooled, high density liquid silica, as modelled by
  the BKS potential.  We find that at density $\rho=4.38$~g/cm$^3$,
  spontaneous nucleation of crystalline stishovite occurs in
  conventional molecular dynamics simulations at temperature
  $T=3000$~K, and we evaluate the nucleation rate $J$ directly at this
  $T$ via ``brute force'' sampling of nucleation events in numerous
  independent runs.  We then use parallel, constrained Monte Carlo
  simulations to evaluate $\Delta G(n)$, the free energy to form a
  crystalline embryo containing $n$ silicon atoms, at $T=3000$,
  $3100$, $3200$ and $3300$~K. By comparing the form of $\Delta G(n)$
  to CNT, we test the ability of CNT to reproduce the observed
  behavior as we approach the regime where spontaneous nucleation
  occurs on simulation time scales.  We find that the prediction of
  CNT for the $n$-dependence of $\Delta G(n)$ fits reasonably well to
  the data at all $T$ studied.  $\Delta \mu$, the chemical potential
  difference between bulk liquid and stishovite, is evaluated as a fit
  parameter in our analysis of the form of $\Delta G(n)$.  Compared to
  directly determined values of $\Delta \mu$ extracted from previous
  work, the fitted values agree only at $T=3300$~K; at lower $T$ the
  fitted values increasingly overestimate $\Delta \mu$ as $T$
  decreases.  We find that $n^\ast$, the size of the critical nucleus,
  is approximately 10 silicon atoms at $T=3300$~K.  At $3000$~K,
  $n^\ast$ decreases to approximately 3, and at such small sizes
  methodological challenges arise in the evaluation of $\Delta G(n)$
  when using standard techniques; indeed even the thermodynamic
  stability of the supercooled liquid comes into question under these
  conditions. We therefore present a modified approach that permits an
  estimation of $\Delta G(n)$ at $3000$~K.  Finally, we directly
  evaluate at $T=3000$~K the kinetic prefactors in the CNT expression
  for $J$, and find physically reasonable values; e.g. the diffusion
  length that Si atoms must travel in order to move from the liquid to
  the crystal embryo is approximately 0.2~nm.  We are thereby able to
  compare the results for $J$ at $3000$~K obtained both directly and
  based on CNT, and find that they agree within an order of magnitude.  In
  sum, our work quantifies how certain predictions of CNT (e.g. for
  $\Delta \mu$) break down in this deeply supercooled limit, while
  others [the $n$-dependence of $\Delta G(n)$] are not as adversely
  affected. 
\end{abstract}

\maketitle
\section{Introduction}

In recent years, computer simulations have increasingly been used 
to study the nucleation and growth of crystals from the supercooled
liquid state.  
Molecular dynamics (MD) simulations have been particularly useful
in testing, on a molecular level, the predictions of classical nucleation
theory (CNT)~\cite{gibbs, volmer, farkas, becker, kelton, pablo}.
%Particularly useful have been the careful tests of
%classical nucleation theory (CNT)~\cite{gibbs, volmer, farkas, becker, 
%kelton, pablo} made possible using the
%molecular-level details of the nucleation process revealed in
%molecular dynamics (MD) simulations~\cite{CCC}.  
In large measure,
this has been made possible by the development of novel computational
techniques that permit the determination, from simulations, of free
energy barriers, kinetic prefactors, and the order parameters required
to quantitatively test the predictions of CNT~\cite{frenkel_1996,frenkel_1999,
frenkel_2001,frenkel_2004,frenkel_2005}.  A key
feature of these techniques is that they allow the study of nucleation
under thermodynamic conditions at which spontaneous crystal nucleation
does not occur on the short physical time scales accessible to
conventional MD simulations.  As a consequence, much previous work has
focussed on testing CNT, and calculating the nulcleation rate, at low
to intermediate degrees of supercooling, where CNT is expected to best
apply.

At the same time, spontaneous crystal nucleation is observed in a
number of simulated liquid systems under very highly supercooled
conditions where the nucleation time is comparable to or less than the
simulation time scale (e.g. Refs.~\cite{md1,md2,md3,md4,md5}). 
Current conventional MD simulations
typically are able to study systems of a few thousand molecules over a
time scale of tens of nanoseconds.  Within these restrictions, a spontaneously
crystallizing system will exhibit quite small crystal nuclei compared
to those found at higher temperature $T$, and it is generally
expected that CNT will not predict well the behavior of the system in
this regime.  Consequently, relatively few studies examine this deeply
supercooled limit of nucleation behavior in the context of CNT.

The purpose of the present work is to explore this deeply supercooled
limit of nucleation behavior, with the goal of testing the limits of
CNT and quantifying how the theory begins to fail in this regime; and
also to determine the technical limits of applicability of the
simulation methods usually employed at higher $T$. We are 
interested in determining if it is possible to compare a nucleation
rate calculated using CNT, and a rate found directly from a
spontaneously crystallizing MD simulation.  The latter question is
particularly interesting, since only a few simulation studies compare
nucleation rates found from CNT to a rate calculated
independently~\cite{frenkel_1999,frenkel_2005,tanaka2005}, 
yet such comparisons are a key tool for
developing and testing improved theoretical descriptions of
nucleation.

To achieve these goals, we study liquid silica as modelled by the BKS
potential~\cite{BKS}.  The thermodynamic and transport properties of the
supercooled liquid state of this model have been characterized in
detail~\cite{horbach,longNAT}.  
Previous work has also evaluated the phase diagram of the
system, providing the coexistence conditions demarcating the liquid,
and several crystalline phases~\cite{PD}.  Most significant for the current
purpose, we find that the liquid spontaneously crystallizes to
stishovite~\cite{stishovite} in our simulations when cooled to approximately $T=3000$~K
at density $\rho=4.38$~g/cm$^3$.  The liquid at this $T$ exhibits the
two-step relaxation in its dynamical quantities characteristic of a
deeply supercooled fluid, but it is still diffusive enough to reach
metastable equilibrium on a time scale much shorter than the time
scale for crystal nucleation.  Consequently, we are able to make a
direct calculation of the rate at $3000$~K using an ensemble of
independent MD simulations, while at the same time, we can determine
the properties of the metastable liquid.

We also use constrained Monte Carlo simulations of the liquid to
calculate the free energy barrier to nucleation at the same density,
over a range of temperatures from $3000$~K to $3300$~K, to test the
degree to which the predictions of CNT are satisfied on approach to
$T=3000$~K.  The key predictions of CNT we wish to test relate to the
central quantity of the theory, $N(n)$, the equilibrium cluster size
distribution,
{%\bf
or the number of clusters containing $n$ particles
}
\cite{pablo}. 
{%\bf
In this work, we will track Si atoms only, and assume from stoichiometry
that a cluster nominally of size $n$ ($n$ Si atoms) actually contains
$3n$ atoms ($n$ Si atoms and $2n$ O atoms).
}
$N(n)$ is interpreted to yield the work $\Delta
G(n)$ of forming a cluster of size $n$ from the surrounding metastable
liquid via,
\begin{equation}\label{eq:delg}
{\boldmath \frac{\Delta G(n)}{k_{\rm B}T}=-\ln{\left[\frac{N(n)}{N(0)}\right]},}
\end{equation}
{%\bf 
where $N(0)$ is the number of liquid-like Si atoms; so defined 
$\Delta G(0)=0$.  Whether the distribution of cluster sizes is extensive
or intensive (i.e., normalized or not), the barrier is system size independent.}  
Within the
CNT framework, the phenomenological model for the work is given by,
\begin{equation}\label{eq:delgCNT}
\Delta G(n) = -\left|\Delta \mu\right| n + a n^{2/3},
\end{equation}
where $\Delta \mu = \mu_{\rm stish} - \mu_{\rm liq}$ is the difference 
in chemical potential between the
bulk stable and metastable phases and $a$ is a surface term that is
proportional to the surface tension $\gamma$ and depends on the shape
of the nuclei.  At a critical cluster size $n^*$, $\Delta G(n)$ has a
maximum and clusters larger than $n^*$ will grow spontaneously,
forming the new phase.  $\Delta G(n^*)$ then represents the free
energy barrier to nucleation. In this study, we use computer
simulation techniques that connect $\Delta G(n)$ with the probability
of appearance of an $n$-sized cluster within the simulation, where the
cluster is identified by a specific cluster criterion~\cite{frenkel_1996,frenkel_2004,richard}. 
We can then compare our barrier calculations with the general form suggested by
Eq. \ref{eq:delgCNT}.

According to CNT, the rate of nucleation, i.e. the rate at which
critical nuclei go over the barrier, is
\begin{equation}\label{eq:JCNT}
J^{CNT}= K \exp{\left(-\frac{\Delta G(n^*)}{k_{\rm B}T}\right)},
\end{equation}
where the kinetic prefactor is given by,
\begin{eqnarray}
K &=& 24 \rho_n Z D {n^*}^{2/3}/\lambda^2 \label{eq:K1} \\
              &=&  \rho_n Z f^+_{\rm crit} \label{eq:K2},
\end{eqnarray}
where $Z=\sqrt{\left|\Delta \mu\right| / 6\pi k_{\rm B}Tn^*}$ is the
Zeldovich factor, $D$ the diffusion constant,
%(in our case of Si atoms)
$k_{\rm B}$ is the Boltzmann constant, $\rho_n$ is the number density of particles,
%(here, Si atoms), 
$\lambda$ is a typical distance particles must diffuse in order to go
from the metastable liquid to the embryonic cluster, and $f^+_{\rm
  crit}$ is the rate at which particles are added to the critical
nucleus.  We note that the use of $f^+_{\rm crit}$ is an innovation
introduced in Ref.~\cite{frenkel_2004}. In the case of diffusive
barrier crossing $f^+_{\rm crit}$ can be calculated from simulation via,
\begin{equation}\label{eq:fcrit}
f^+_{\rm crit} = \frac{1}{2}\frac{\left<\left[n^*(t)-n^*(0) \right]^2 \right>}{t},
\end{equation}
where $\left<.\right>$ denotes an ensemble average.

Our Monte Carlo simulations of liquid silica between $3300$ and
$3000$~K show that CNT describes the liquid well at the highest $T$,
but that deviations in the observed and predicted behavior emerge at
lower $T$.  At the lowest $T$, we also identify technical difficulties
associated with obtaining $N(n)$ and we describe an alternative
strategy that at least partially addresses them.  Notwithstanding these
challenges, at the lowest $T=3000$~K, we are still able to calculate
the kinetic prefactors for the nucleation rate as described in CNT, so
that we can compare the predicted rate to that calculated from direct
MD simulations.  Despite the worsening correspondence between our
results and the thermodynamic aspects of CNT at low $T$, the rates
compare reasonably well.  Whether the correspondence of the rates at
this large degree of supercooling is peculiar to our system, or
whether this is a general result is an open question. We also find
that $N(n)$, as obtained for the equilibrium system
(i.e., the system that samples the equilibrium distribution of embryos,
including those embryos near to, at, and beyond the nucleation barrier), 
is different from the analogous quantity for the metastable liquid state
(i.e., the metastable equilibrium sampled in a conventional MD
simulation prior to the onset of nucleation), raising the
question of which distribution is more significant in determining the
rate.

In Section II, we desscribe the model system.  Section III describes the
direct MD nucleation rate calculation, while Section IV describes the
CNT calculations and explores the use of the metastable
liquid in determining the free energy barrier.  We subsequently
present our Discussion and Conclusions.  Appendix I describes our
methods and criteria for defining a crystalline cluster.

\section{System of study}

\begin{figure}
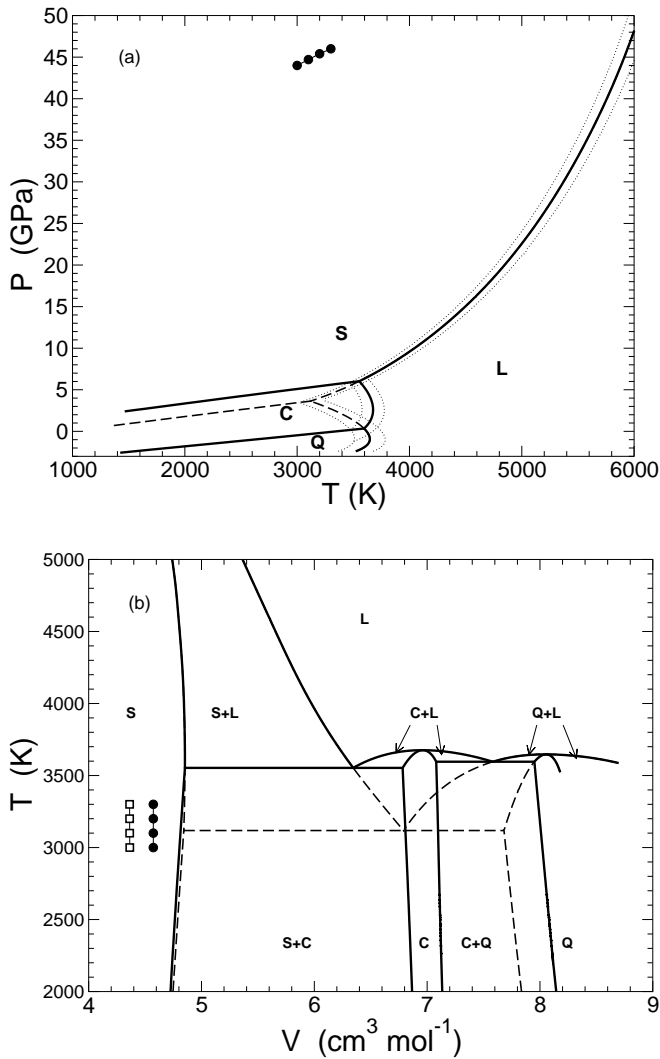

\hbox to \hsize{\epsfxsize=1.0\hsize\hfil\epsfbox{fig1a.eps}}
\vspace{0.5cm}
\hbox to \hsize{\epsfxsize=1.0\hsize\hfil\epsfbox{fig1b.eps}}
\caption{Location of studied liquid state points (filled circles) in
  the (a) $PT$ and (b) $VT$ phase diagram of BKS silica.  At a given
  $P$ and $T$, stishovite has a higher density than that of the
  liquid.  Therefore, we also plot the stishovite state points
  corresponding to the liquid $P$ (open squares) in (b).  The phases
  shown in the diagrams are the liquid (L), $\beta$-quartz (Q),
  coesite (C) and stishovite (S).  (b) At fixed $V$, thermodynamic
  ground states are often mixtures of two coexisting phases.  The
  liquid state points studied fall within
  the one-phase stability field of stishovite.  Despite
  the proximity of our chosen state points to the stishovite to
  stishovite-plus-coesite boundary in the $VT$ plane, the location in
  the $PT$ plane is deep within the stishovite region.  Dashed lines
  are metastable extensions of stishovite and $\beta$-quartz
  transitions, and dotted lines represent the uncertainty in the
  location of the melting lines.  $V$ is given per mol ion.
 }
\label{fig:pd}
\end{figure}

\begin{figure}
\hbox to \hsize{\epsfxsize=1.0\hsize\hfil\epsfbox{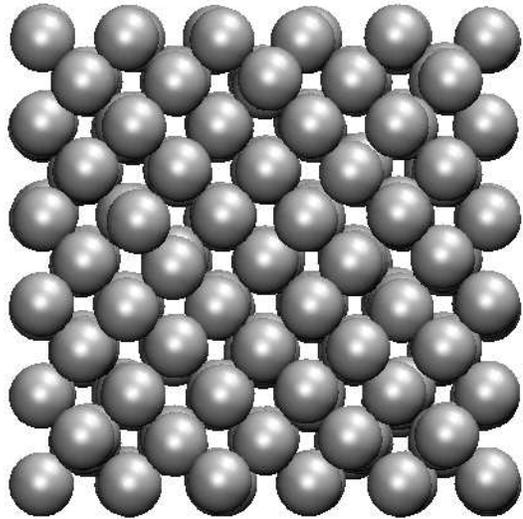}}
\caption{
Stishovite at $3000$~K, as viewed down the crystallographic
$c$-axis. Only Si atoms are shown.  
 }
\label{fig:stish}
\end{figure}

We study a system of $444$ Si and $888$ O ions governed by a modified
BKS potential and in a cubic simulation cell with periodic boundaries.  
The modification includes
a short range additive patch to prevent ``fusion'' events, and a
tapering of the real space part of the potential with a polynomial
tail so that it smoothly reaches zero at $1$~{nm}.  Long range forces
are handled via the Ewald summation.  The details of the potential are
given in Ref.~\cite{longNAT}.

The phase diagram of BKS silica has been recently evaluated in
Ref.~\cite{PD} (Fig.~\ref{fig:pd}).  In that work, the stability
fields of the liquid, stishovite, $\beta$-quartz and
coesite have been determined. In this
study we focus on the molar volume $V=V_0=4.5733$~cm$^3$/mol
($\rho=4.3793$~{g cm$^{-3}$}) liquid isochore, 
which as shown in Fig.~\ref{fig:pd}(b),
falls in the one-phase stability field of stishovite.  A view of
stishovite along the $c$-axis showing Si atoms only is provided in
Fig.~\ref{fig:stish}.  In stishovite, there are six O atoms
surrounding each Si atom in an octahedral arrangement.  These
octahedra, connected along edges and at corners, arrange themselves in
a compact manner.

We perform our simulations in the $NVT$ ensemble, where $N$ is the 
number of molecules.  Usually, nucleation
experiments and simulations are done at constant pressure, $P$.
However, we find that for the state points of interest, the critical
nuclei we observe are small, and do not noticeably change the $P$ of
the system.  Only once the crystallization process advances well into
the growth stage does $P$ or the potential energy $U$ of the system
change significantly.  Therefore, the critical nucleus forms within a
liquid that is characterized to a good approximation either by the
system's $P$ or $V$.  However, the density of the nucleus itself is
not known.  Hence, we also show in Fig.~\ref{fig:pd}(b) the $V$ 
of bulk stishovite at the pressure at which the liquid is studied.

$P$ and $U$ are needed along the $V_0$ isochore to determine
$\Delta \mu$, which we obtain by extending the calculations  
described in Ref.~\cite{PD}.
%  where $\Delta \mu$ between the two phases was calculated at lower $P$.} [?????] 
The diffusion coefficient is also needed in order to calculate the kinetic
prefactor. To obtain these quantities, we perform MD simulations at
constant $V$ in both the liquid and stishovite, near and along $V_0$.
First we equilibrate the system near the desired $T$ with simple
velocity scaling every 100 timesteps, and then we allow the system to
continue in the $NVE$ ensemble for about 1~ns ($E$ is the total energy).  
For the $T=3000$~K and
$T=3100$~K cases, numerous independent runs are performed, and only
those that do not show any signs of crystallizing are used to
determine desired quantities.

For the liquid at $T=T_0=3000$~K and $V_0$, the Si diffusion coefficient
is $D=8.04\pm0.2 \times 10^{-7}$~{cm$^2$~s$^{-1}$} as determined from
the slope of the mean squared displacement of Si atoms as a function
of time $t$, and the pressure is found to be $P_0 = 44.0$~GPa.  This and
higher $T$ state points are shown to be in the stishovite stability
field in the $PT$ phase diagram [Fig.~\ref{fig:pd}(a)].

At $P_0$, stishovite has a lower molar volume
($V=4.364$~cm$^3$~mol$^{-1}$) than the melt.  The stishovite state
points corresponding to the pressure of the liquid  
[shown in Fig.~\ref{fig:pd}(b)]   are the state points used to
calculate $\Delta \mu$: $\Delta \mu$ is calculated between the two
phases at the same $P$, for which (in general) the volumes are
different.

%\subsection{Molecular dynamics simulation}
\section{Nucleation rate from MD simulations}

\begin{figure}
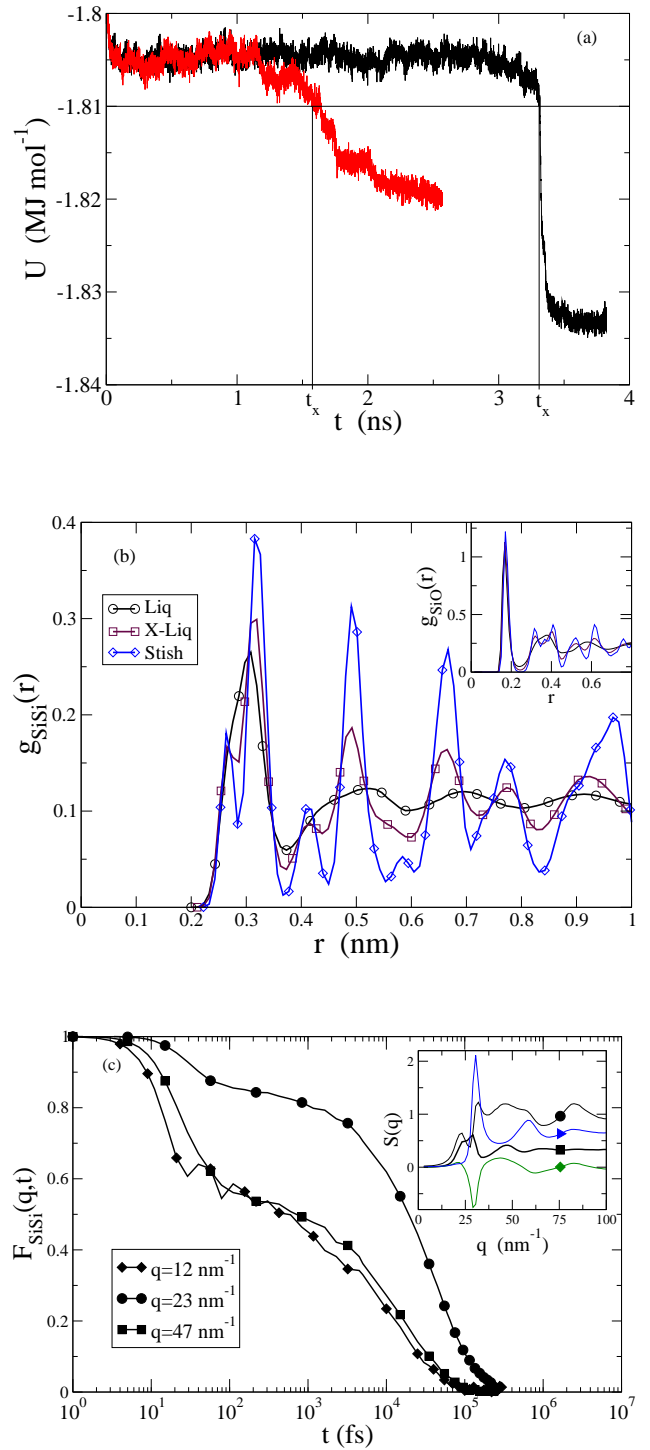

\hbox to \hsize{\epsfxsize=0.96\hsize\hfil\epsfbox{fig3a.eps}}
\vspace{1.0cm}
\hbox to \hsize{\epsfxsize=0.96\hsize\hfil\epsfbox{fig3b.eps}}
\vspace{0.9cm}
\hbox to \hsize{\epsfxsize=0.96\hsize\hfil\epsfbox{fig3c.eps}}
\caption{
Crystalizing liquid.  (a) Determination of $t_x$ from the
potential energy as a function of time. At $t=0$, the thermostat is
reset from $5000$~K to $3000$~K.  When the system reaches a
potential energy of $U_x=-1.81$~(MJ/mol), it is well underway to
crystallizing.  (b) Structure at $3000$~K as measured by $g_{\rm
SiSi}(r)$ and $g_{\rm SiO}(r)$ (inset), of the liquid (Liq), the
system after crystallization (X-Liq) and stishovite (Stish).  (c)
$F_{\rm SiSi}(q,t)$, the dynamic structure factor for Si atoms for
three $q$.  Inset shows the static structure factor $S(q)$ along
with its components: total (circle), SiSi (square), OO (triangle) ,
and SiO (diamond).  
}
\label{fig:tx}
\end{figure}

We use MD simulations in the $NVT$ ensemble to calculate the
nucleation rate at $T_0$ and $V_0$ in a ``brute force'' way.  We first
equilibrate the liquid at $5000$~K, and then quench $198$ independent
configurations to $3000$~K, employing the Berendsen
thermostat~\cite{BERENDSEN} with a time constant of $1$~ps.  Other
simulation details are the same as given in the previous section.  The
simulation continues at $3000$~K until the system crystallizes.

Fig.~\ref{fig:tx}(a) shows two sample time series, illustrating a
large drop in potential energy associated with the phase change.
Fig.~\ref{fig:tx}(b) shows components of the radial distibution
function $g(r)=g_{\rm SiSi}(r) + 2 g_{\rm SiO}(r) + g_{\rm OO}(r)$ for
the metastable liquid (i.e., when the time series is stable), the
crystallized system, and pure stishovite at $T_0$.  The comparison of
each $g(r)$ shows that we indeed crystallize to stishovite.

Fig.~\ref{fig:tx}(c) shows $F_{\rm SiSi}(q,t)$, the dynamic structure
factor at fixed wavenumber $q$ obtained by considering only Si atoms (which diffuse more
slowly than O), obtained from the metastable liquid portion of a
simulation before the onset of crystallization.  As a reference for
the three wavenumbers chosen, we plot the static structure factor
$S(q)$ in the inset.  We do not observe any time evolution of $S(q)$
during the steady state liquid portions of the time series.  With
regards to the $NVE$ simulations, we do not observe any significant
differences in $F(q,t)$ or $S(q)$.

From $F_{\rm SiSi}(q,t)$, we see that the $\alpha$-relaxation time for
the system at $V_0$ and $T_0$ is approximately $100$~ps.  Thus, even
though the system exhibits two-step (glassy) relaxation, the
relaxation time is typically much shorter than the nucleation times
and we are able to achieve a metastable liquid state.
%, despite being deeply supercooled.

If the fraction $R(t)$ of unnucleated systems obeys a simple
first-order rate law, then the rate of nucleation $J$ can be obtained
from
\begin{equation}\label{eq:rate}
\ln{[R(t)]}=-JV \left(t-t_0\right)\mbox{ ,}
\label{rate}
\end{equation}
where $V$ is the volume of the system and $t_{0}$ is the lag time,
i.e. the time required to achieve a steady state of precritical
nuclei~\cite{bartell_1999}. However, to take advantage of
Eq.~\ref{rate} we must be able to identify the time when a particular
system from our ensemble of runs has nucleated. Various methods can be
used to detect crystallization such as examing the Voronoi volumes or
counting the number of particle neighbors~\cite{bartell_1999}.  In
this study, we contrast three approaches: first, we employ a simple
energy criterion specific to the state point studied so that the
crystallization time $t_x$ for a MD run is the time at which the
potential energy first reaches a value of $U_x=-1.81$~{MJ/mol}.  That
very few simulation runs reached $U_x$ and then returned to a steady
state liquid means that the system has progressed well past
nucleation. In a sense, this mimics experimental measures which only
identify a nucleation event by observing post critical clusters that
are growing. Second, we allow runs to continue $500$~ps
after the system reaches a lower potential energy $U_{\rm
  low}=-1.82$~MJ/mol. Using $U_{\rm low}$, we can also check the
sensitivity of our rate calculation on the chosen energy threshold.

Our third criterion is based on identifying the critical nucleus. In
Section IV, we find the size of the critical nucleus, using the
cluster criteria outlined in Appendix I to identify an $n$-sized
cluster, to be about $3$. We can then define a new time, $t_{\rm nuc}$ as
the latest time at which the largest cluster in the system $n_{\rm
  max}\le 1$, i.e., the last time the liquid is precritical, and
compare the nucleation rates based on our different criteria.
By choosing $n_{\rm max}\le 1$, we err on the side of making $t_{\rm nuc}$
a lower bound on the nucleation time.

{%\bf
Thus, we have three measures of the nucleation time:  the time
$t_x$ it takes to reach a potential energy $U_{\rm x}$ indicative
of the beginning of crystallization; the time
it takes to reach a low energy threshold $U_{\rm low}$;
and the last time $t_{\rm nuc}$ at which the system possesses a
largest cluster of size $1$.
}

\begin{figure}
\hbox to \hsize{\epsfxsize=1.0\hsize\hfil\epsfbox{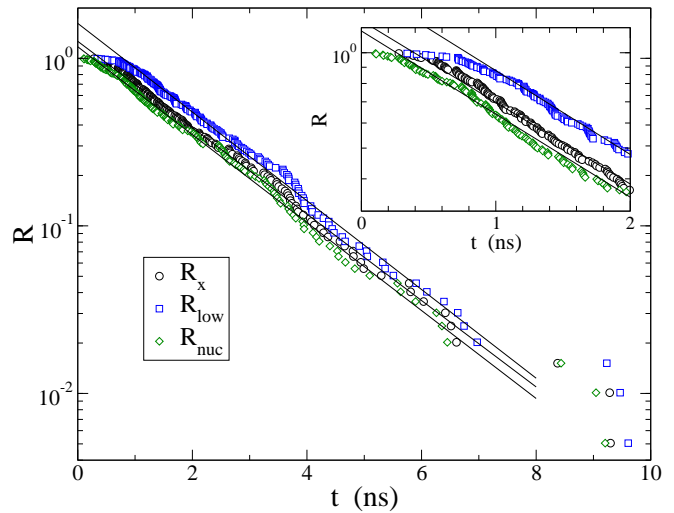}}
\caption{
Determination of nulceation rate from $R$.  The function
$R_{\rm x}$ (circles) is shown here to be well described (except for
very early times) by exponential decay with rate constant $0.60 \pm
0.02$~{ns}$^{-1}$, determined from the line of best fit.  Data
between 1~ns and 5~ns were used to obtain the fit.  The functions
$R_{\rm low}(t)$ and $R_{\rm nuc}(t)$ have the same slope to within
uncertainty.  Inset: close-up at small $t$.
}
\label{fig:rate}
\end{figure}

Fig.~\ref{fig:rate} shows a plot of $R$ using these three criteria
$R_{\rm x}$, $R_{\rm low}(t)$ and $R_{\rm nuc}(t)$, 
obtained using the upper and lower energy criteria
and the critical cluster criteria respectively, as a function of
time. The slope of each of these functions is the same within the
error suggesting that our rate calculation is not highly sensitive to
the nucleation criteria, while the lag time, obtained from the
intercept, is sensitive.  Using $t_x$, we find that the $198$ quenches
from $5000$~K to $3000$~K yield a shortest crystallization time of
$0.28$~ns, and longest time of $10.53$~ns, with an average time of
$2.12$~ns.  The slope of the line of best fit is $0.60 \pm
0.02$~{ns}$^{-1}$, where we have omitted data from times before
$1$~ns in the fit, and the time lag $t_0=0.4$~ns. Given that our
cubic box length is $2.1627$~nm, this yields a rate of $J=0.059 \pm
0.002$~{nm}$^{-3}${ns}$^{-1}$, or $6 \times
10^{34}$~{m}$^{-3}${s}$^{-1}$.

%{\bf UNITS ps to ns}

We measure $t_{\rm nuc}$ in each of our $198$ nucleation runs to a
resolution of $0.005$~ns and find a lower estimate of the lag time
$t_0\approx0.260$~ns. The average time difference is $t_x - t_{\rm nuc}=
0.212$~ns, with standard deviation $0.117$~ns. Comparing these two criteria,
we find the mean value of the largest cluster at $t_x$ is $\left<
  n_{\rm max}(t_x)\right>=39$, i.e. about $10\%$ of the system, with a
standard deviation of $23$.  Therefore, for our system, the
crystallization process significantly lowers the energy only when
about $10\%$ of the system has crystallized. We also note that
in about $5\%$ of the runs, the $U_x$ criterion is triggered
prematurely, i.e., a low-energy fluctuation goes below $U_x$, but then
the system energy remains in steady state.

We note that calculating the rate directly from the slope of
the plots in Fig.~\ref{fig:rate} may ignore effects due to transient
nucleation, i.e. that $R$ decreases smoothly at early times
instead of remaining at $1$ until $t_0$~\cite{bartell_1999}, so that our
estimate of the rate and lag times are strictly lower
bounds. Nevertheless, the independence of our slope on the nucleation
criteria suggests that our calculation of the rate is robust and any
corrections would be small.

\begin{figure}
\hbox to \hsize{\epsfxsize=1.0\hsize\hfil\epsfbox{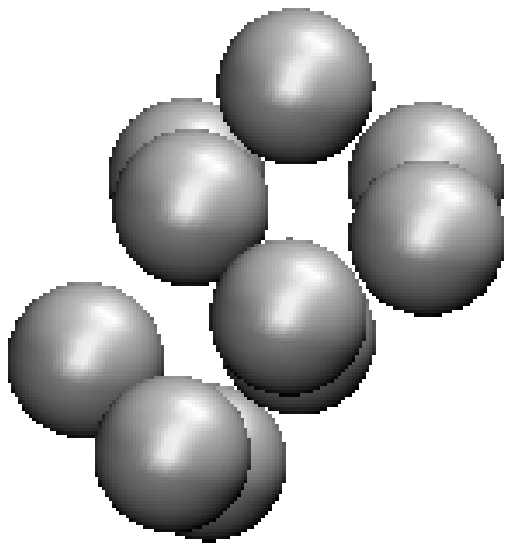}}
\hbox to \hsize{\epsfxsize=1.0\hsize\hfil\epsfbox{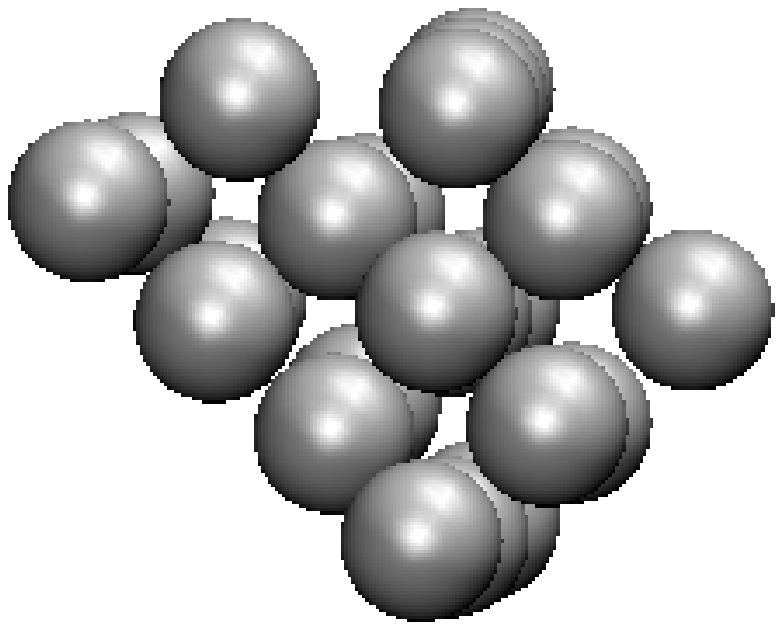}}
\hbox to \hsize{\epsfxsize=1.0\hsize\hfil\epsfbox{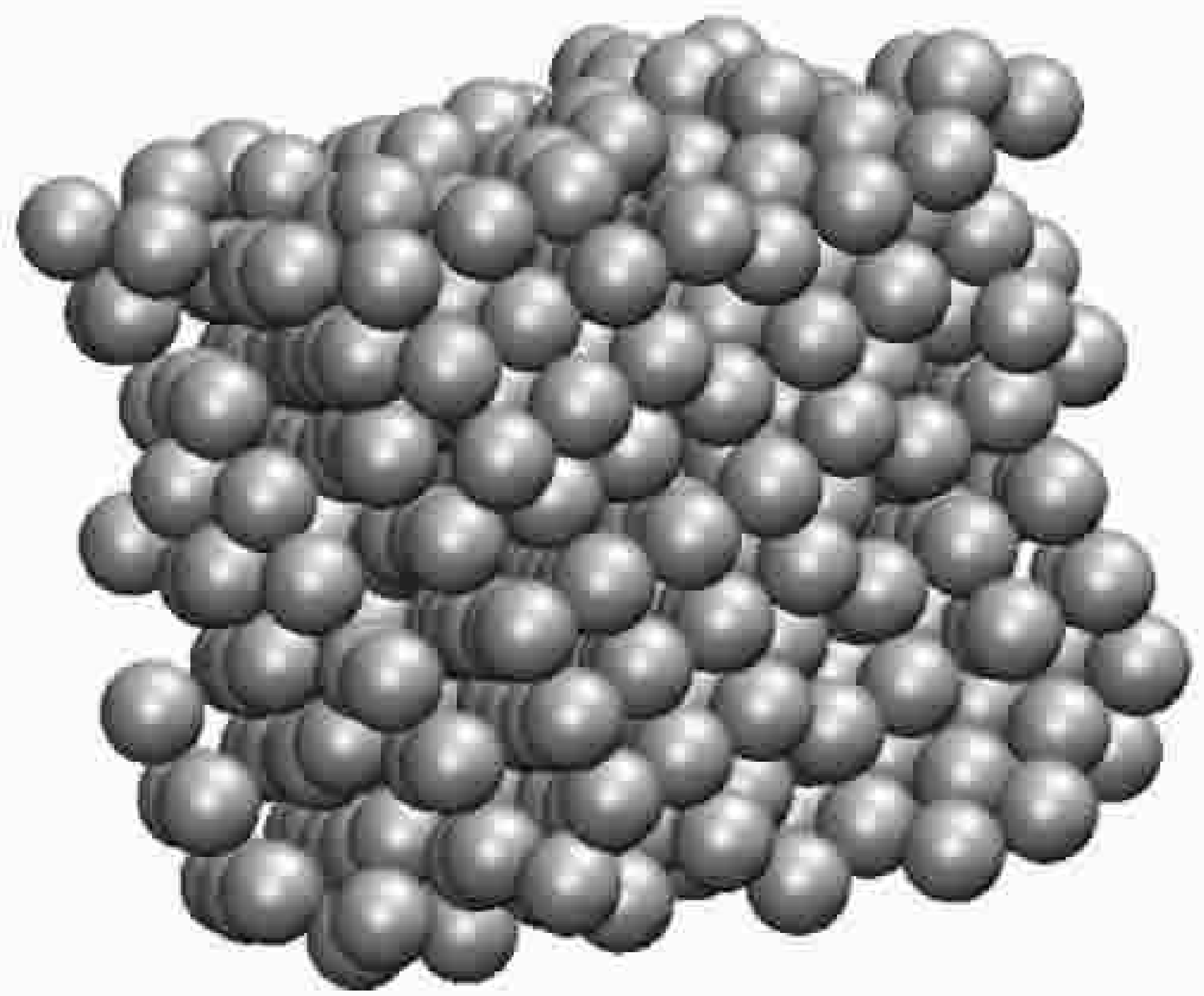}}
\caption{
Top: Sample critical nucleus at $3300$~K containing $10$ Si
atoms.  Middle: A snapshot of the growing crystal embryo from a
dynamic crystallization simulation at $3000$~K when it contains 23
Si atoms.  Bottom: Sample end configuration of a crystallization
simulation. 
}
\label{fig:nice}
\end{figure}

Fig.~\ref{fig:nice}(a) shows a critical nucleus at $3300$~K (obtained
from constrained MC simulations described later), while
Fig.~\ref{fig:nice}(b) shows a postcritical crystallite of size $23$
from an MD simulation at $T_0$.  Fig.~\ref{fig:nice}(c) shows a
representative configuration at the end of a crystallization
simulation.  These pictures provide a visual confirmation that the
liquid does indeed crystallize to stishovite; that small nuclei (at least for 
$n_{\rm max}\ge 10$)
resemble the bulk phase; and that the procedure used to define
clusters is able to track the nucleation process.

\section{CNT calculations}

\subsection{Free energy barrier}

The central quantity of CNT is $N(n)$.  However, it is not feasible to
obtain $N(n)$ through direct simulation for two reasons: critical
clusters are typically rare, and hence it is difficult to gain
statistics; and after the formation of the critical cluster, the
system irreversibly evolves toward the crystalline state. To overcome this,
we add a bias to the system Hamiltonian which constrains the clusters
of interest into existence. The new constrained Hamiltonian is then
\begin{equation}\label{eq:Hc}
H_{\rm C}= H_{\rm BKS} + \phi(n_{\rm max}),
\end{equation} 
where $H_{BKS}$ is the unbiased Hamiltonian derived from our BKS
potential and
\begin{equation}
\phi(n_{\rm max})= \frac{\kappa}{2}\left( n_{\rm max} - n_0 \right)^2\mbox{ ,}\\
\end{equation}
is the contraint with $\kappa$ and $n_0$ being constants, and where
$n_{\rm max}$, the size of the largest cluster in the system, is an
order parameter~\cite{frenkel_2004}. The $N(n)$ measured under the
constraint is then related to its value in the unconstrained system
through the relation
\begin{equation}\label{eq:convert}
  \left<N(n)\right>=\frac{ \left< N(n) \exp{\left[\phi(n_{\rm max})/k_{\rm B}T\right]}\right>_C }{\left<\exp{\left[\phi(n_{\rm max})/k_{\rm B}T\right]}\right>_C},
\end{equation}
where $\left<.\right>_C$ denotes an average in the constrained
ensemble. In the case where a cluster of size $n$ is rare
$N(n)=P(n_{\rm max})$, the probability that the largest cluster in the 
system is of size $n_{\rm max}$, and Eq.~\ref{eq:convert}
becomes,
\begin{equation}\label{eq:convert2}
  \left<P(n_{\rm max})\right>=\frac{ \left< P(n_{\rm max}) \right>_C \exp{\left[\phi(n_{\rm max})/k_{\rm B}T\right]}}{\left<\exp{\left[\phi(n_{\rm max}) /k_{\rm B}T\right]}\right>_C}.
\end{equation}
It is important to note that our cluster definition ignores O atoms,
and only uses Si atoms.  Thus, a cluster of size $n$ contains $n$ Si
atoms, or $n$ SiO$_2$ units (see Appendix). 

Since it is easier in practice to measure $P(n_{\rm max})$ than $N(n)$,
we will use $P(n_{\rm max})$ interchangeably with $N(n)$ in the regime where 
the two are shown to be equal.  Formally, this occurs when clusters are rare,
and can be justified by the following.  Let $P_n$ be the probability that there is
at least one cluster of size $n$ in the system, and $P_n(i)$ be the probability that
there are exactly $i$ clusters of size $n$.  Then,
\begin{eqnarray}
P_n &=& P_n(1) + P_n(2) + P_n(3) \dots \\
N(n) &=& P_n(1) + 2P_n(2) + 3P_n(3) \dots.
\end{eqnarray}
What we mean by a rare cluster of size $r$ is that $P_r(1)$ is small, and additionally
that rare cluster appearance is independent of what other clusters are present, i.e.
$P_r(2)\approx P_r(1)\times P_r(1) \approx 0$.
This immediately leads to $P_n=N(n)$~\cite{richard,frenkel_2004}.
By extension, two rare clusters of different sizes appearing at the same time also occurs 
with vanishing probability $P_{r+m}(1)\times P_r(1) \approx 0$ 
[assuming $P_{r+m}(1) < P_{r}$ for $m>0$, i.e. larger clusters are rarer], 
and so a rare cluster will also be the largest cluster in the system.  
From these arguments, we obtain $P_n(1) = P_n = N(n) = P(n_{\rm max})$, for $n\ge r$
(the equality holds up to a normalization constant that is irrelevant in determining
the free energy).  
Of course, when $N(n) \ne P(n_{\rm max})$, we measure $N(n)$ directly.

The basic MC scheme follows that presented in Ref.~\cite{hybrid},
where short $NVE$ MD trajectories generate new configurations that are
tested against the Boltzmann distribution.  More explicitly, we begin
with a configuration $C_1$ with largest cluster $n_{\rm max}^{[1]}$.
New random velocities drawn from the Maxwell distribution appropriate
to the desired $T$ are assigned to all the particles, and at this
point the total energy is $H_{\rm BKS}^{[1]}$.  With these new
velocities, the system evolves along a constant NVE MD trajectory for
10 timesteps, with forces derived from $H_{\rm BKS}$, to arrive at a
new configuration $C_2$ with total energy $H_{\rm BKS}^{[2]}$, and
largest cluster $n_{\rm max}^{[2]}$.  With a perfect integration
scheme, $H_{\rm BKS}^{[2]}=H_{\rm BKS}^{[1]}$.  $C_2$ is accepted with
propability $p$ given by,
\begin{equation}
p =  \min \left \{1, \exp \left[  -\frac{1}{k_{\rm B} T} \left( H_{\rm C}^{[2]} - H_{\rm C}^{[1]} \right)\right] \right \}.
\end{equation}
It is important to note that the acceptance criterion for this hybrid
MD-MC method uses the total energy (kinetic plus potential), rather
than just the potential.

With this hybrid MC method, it is only necessary to evaluate the
cluster size distribution of the system at the end of each MD
mini-trajectory.  The method also provides a way of incorporating the
Ewald sums through multiparticle MD moves, i.e., energy changes
arising from single particle moves are difficult to calculate
efficiently when there are long range forces.

In order to facilitate equilibration, we employ parallel tempering
over a matrix of runs having different values of $n_0$ and $T$.  Our
tempering scheme follows the descriptions given in 
Refs.~\cite{frenkel_2004,frenkel_book}.
The particulars are as follows.  Compute nodes running in parallel decide
whether to attempt switches of configurations with neighboring nodes
every 10 MC steps, alternating between $T$-switch and $n_0$-switch
attempts.  For $T$-switches, an attempt is made with each neighbor
with probability $0.36$.  For $n_0$-switches, an attempt is made with
each neighbor with probability $0.19$.  The probabilities for
accepting switches are given in Refs.~\cite{frenkel_2004,frenkel_book}.
{%\bf
In practice, it is computationally faster to switch Hamiltonians or $T$
between processors, rather than configurations.
}

\begin{table}
\begin{tabular}{|c|c|c|c|c|c|c|c|c|c|c|}
\hline
 (a) &  \multicolumn{10}{|c|}{$n_0$} \\
 \hline
T  & 1  & 3 & 5 &  7  & 9  & 11 & 14 & 17 & 20 & 25 \\
\hline
3000  &  0  &  1  &  2  &  3  &  4  & & & & & \\
3100  &  5  &  6  &  7  &  8  &  9  &  10  &  11 & 12 & & \\
3200  &  13 & 14 & 15 & 16 & 17 & 18 & 19 & 20 & 21 & 22 \\
3300  &   23 &  &  24 & & 25 & & 26 & 27 & 28 & 29 \\
\hline
%\multicolumn{11}{|c|}{} \\
%\hline
(b)  &  \multicolumn{10}{|c|}{} \\
\hline
3100  &  0  &  1  &  2  &  3  &  4  & & & &  \multicolumn{2}{|c|}{}\\
3200  &  5  &  6  &  7  &  8  &  9  &  10  &  11 & 12 & \multicolumn{2}{|c|}{}\\
3300  &  13 & 14 & 15 & 16 & 17 & 18 & 19 & 20 & \multicolumn{2}{|c|}{} \\
\hline
%\multicolumn{11}{|c|}{} \\
%\hline
(c) &  \multicolumn{10}{|c|}{} \\
\hline
3000  &  0  &  1  &  2 &  \multicolumn{7}{|c|}{}\\
3100  &  3  &  4  &  5 & \multicolumn{7}{|c|}{} \\
3200  &  6 & 7 & 8 & \multicolumn{7}{|c|}{}\\
\hline
\end{tabular}
\caption{Parallel tempering grids for three sets of simulations showing $n_0$ and $T$ for each node, using a parabolic constraint on $n_{\rm max}$ with $\kappa=8$~kJ/mol (a), and $\kappa=16$~kJ/mol (b-c).  Each node is allowed to communicate with its nearest neighbor.  For example, in (b), node 10 can attempt $T$-tempering switches with node 18, and $n_0$-tempering switch with nodes 9 and 11.}
\label{tab:sim2}
\end{table}

To gather data, we set up a grid of simulations with several values of
$n_0$ for each $T$.  Simulations are seeded from configurations
sampled from the MD crystallization runs at $T_0$.  To initially
locate $n^*$, we set up a grid as given in Table~\ref{tab:sim2}(a),
with $\kappa=8$~{kJ/mol}.  After equilibration, and after roughly
determining the shape of the $\Delta G(n)$ curves, we set up other
grids with $\kappa=16$~{kJ/mol}, as shown in
Table~\ref{tab:sim2}(b-c).

Our results for $T=3100$, $3200$ and $3300$~K come from
Table~\ref{tab:sim2}(b), and the run time is $700\,000$ MC steps.  The
starting configurations are those from Table~\ref{tab:sim2}(a).
$\Delta G(n)$ is calculated over the interval from $100\,000$ to
$700\,000$ MC steps, checking that $\Delta G(n)$ as calculated
separately from the intervals $100\,000-400\,000$ MC steps and
$400\,000-700\,000$ MC steps do not show appreciable differences.  For
these $T$, $N(n)\approx P(n_{\rm max})$ for $n \ge 2$ and so we use
$P(n_{\rm max})$ to determine $\Delta G(n)$.  The first part of
$\Delta G(n)$ is obtained by calculating $N(n)$ directly from
simulations where $n_0$ is small. The simulation grid in
Table~\ref{tab:sim2}(c) provides a consistency check on the results
for $T=3100$~K and $T=3200$~K.  The run time is $350\,000$ MC steps.

\begin{figure}
\hbox to \hsize{\epsfxsize=1.0\hsize\hfil\epsfbox{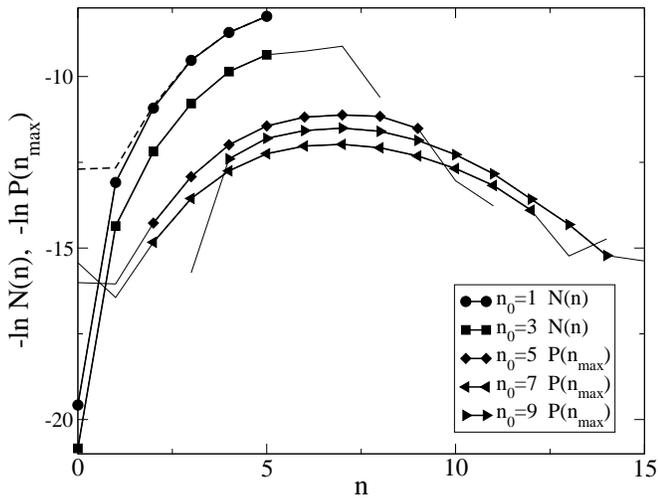}}
\caption{
Portions of $\Delta G(n)$ before shifting, based on
unnormalized histograms for $N(n)$ and $P(n_{\rm max})$.  These are data
transformed via Eq.~\ref{eq:convert} and~\ref{eq:convert2} from the
constrained into the BKS silica ensemble.  These distributions are
for $T=3200$~K, for processors 5, 6, 7, 8 and 9 from
Table~\ref{tab:sim2}(b).  Legend shows the values of $n_0$ used to
constrain the system.  Symbols and bold lines indicate portions of
the data used to obtain the complete $\Delta G(n)$.  For the cases
of $n_0=1$ and $3$, $N(n)$ is obtained directly.  The dashed line
shows $P(n_{\rm max})$ for the $n_0=1$ case, illustrating that
already  $N(n)=P(n_{\rm max})$ for $n\ge 2$.  For larger values of $n_0$,
$P(n_{\rm max})$ is used to obtain $\Delta G(n)$.
}
\label{fig:show3200}
\end{figure}

\begin{figure}
\hbox to \hsize{\epsfxsize=1.0\hsize\hfil\epsfbox{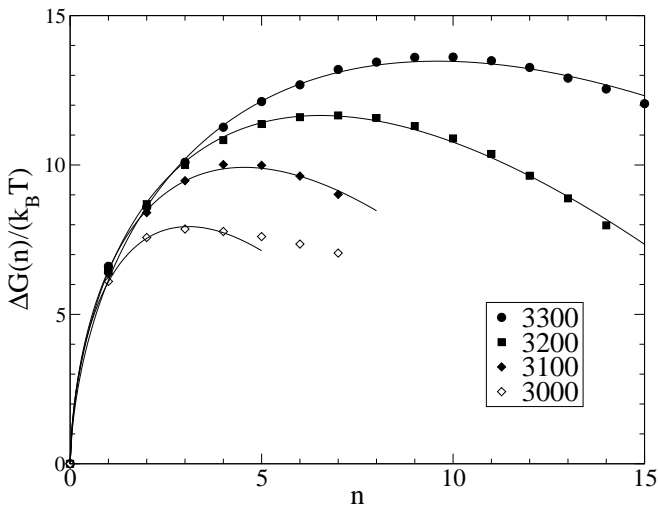}}
\caption{
$\Delta G(n)$ obtained from $N(n)$ after piecing together
results from parallel simulations such as those shown in
Fig.~\ref{fig:show3200}.  Filled symbols are for data described in
Table~\ref{tab:sim2}(b), open diamonds are for hard wall constraints
described in Table~\ref{tab:sim3}.  The solid curves are fits to the
form given by Eq.~\ref{eq:delgCNT}.  For $T=3000$~K only points with
$n\le 4$ are used for the fit.  For $T=3300$, a one-parameter fit is
shown: $\Delta \mu$ in Eq.~\ref{eq:delgCNT} is obtained from
independent calculations, and only $a_{\rm fit}$ is left to fit.
}
\label{fig:profiles}
\end{figure}

Fig.~\ref{fig:show3200} shows pieces of $N(n)$ and $P(n_{\rm max})$
obtained from parallel simulations for $T=3200$~K and for various
$n_0$.  We see from the $n_0=1$ case that for $n \ge 2$, $P(n_{\rm
  max})=N(n)$.  Therefore, only $P(n_{\rm max})$ need be calculated
for determining $\Delta G(n)$ beyond $n=2$.  Fig.~\ref{fig:show3200}
also shows the consistency of the sampling between simulations of
different $n_0$ near the top of the barrier.  For each $n_0$, a
portion of $N(n)$ is recovered up to a multiplicative constant, or
additive constant in $\Delta G(n)$.  The pieces are matched using the
self-consistent histogram method~\cite{frenkel_book}, and the
resulting $\Delta G(n)$ curve is shown in Fig.~\ref{fig:profiles}.
The curves for $T=3300$ and $3100$~K are produced by the same method.

\subsection{Methodological challenges at $T=3000$~K}

For $T=3000$~K, we encounter methodological difficulties
using the parabolic constraint in our MC simulations, apparently due
to the small size of the critical nucleus.  At this $T$,
as we shall see, $P(n_{\rm max})\ne N(n)$, and so we must find $N(n)$
directly, but the parabolic constraint together with
Eq.~\ref{eq:convert} does not yield adequate statistics.  For example,
in Table~\ref{tab:sim2}(c), node $0$ infrequently samples states over
the barrier, and because of the large factor of
$\exp{\left[\phi(n_{\rm max})/k_{\rm B}T\right]}$ in
Eq.~\ref{eq:convert}, these particular states dominate the resulting
$N(n)$.

\begin{table}
\begin{tabular}{|c|c|c|c|c|c|c|}
\hline
  &  \multicolumn{6}{|c|}{$n_{\rm max}^{l} - n_{\rm max}^{u}$} \\
\hline
T & 0-2 & 1-3 & 2-4 & 3-5  & 4-6 &  5-7  \\
\hline
3000  &  0  &  1  &  2  & 3 & 4 & 5\\
3100  &  6  &  7  &  8 & 9 & 10 & 11 \\
\hline
\end{tabular}
\caption{Parallel tempering simulation grid showing $T$ and limits on $n_{\rm max}$ for each node, using hard wall constraints.  For a given node, only configurations with $n_{\rm max}^l \le n_{\rm max} \le n_{\rm max}^u$ are accepted during the MC simulation.  Each node is allowed to communicate with its nearest neighbor.  For example, node 1 can attempt $T$-tempering switches with node 7, and $n_{\rm max}$-tempering switches with nodes 0 and 2.}
\label{tab:sim3}
\end{table}

To obtain $N(n)$ at $3000$~K, we replace the parabolic constraint with
a vertical, hard wall potential by setting upper and lower bounds on
$n_{\rm max}$.  Any MC move which violates $n^l_{\rm max} \le n_{\rm
  max} \le n^u_{\rm max}$ is rejected.  Eq.~\ref{eq:convert} is
modified to simply be $\left<N(n)\right> = \left<N(n)\right>_C$ for
$n^l_{\rm max} \le n_{\rm max} \le n^u_{\rm max}$,
{%\bf 
i.e., $N(n)$ is the same in the constrained and unconstrained ensembles
within the upper and lower bounds on $n_{\rm max}$.}
The constrained
Hamiltonian in this case becomes,
\begin{equation}\label{eq:HW}
H_C = H_{BKS} + \delta(n_{\rm max};n^l_{\rm max},n^u_{\rm max}),
\end{equation}
where
\begin{eqnarray}
\delta(n_{\rm max}) = &0& {\rm for} \,\, n^l_{\rm max} \le n_{\rm max} \le n^u_{\rm max}  \\
              &\infty & {\rm otherwise.}
\end{eqnarray} 
We then set up the hard wall simulation as outlined in
Table~\ref{tab:sim3}, taking initial configurations from
Table~\ref{tab:sim2}(c) and running for $1\,400\,000$ MC steps.  For each
node, $N(n)$ is determined for $n^l_{\rm max} \le n \le n^u_{\rm
  max}$.  Small windows in $n$ are use to gather good statistics as
well as to prevent nucleation in the bin closest to $n=0$.  Error
estimates are taken by considering different time intervals in
determining $N(n)$.  Ideally, $\Delta G(n)$ for $T=3100$~K should be the
same when calculated either with hard wall or parabolic constraints.
Any discrepancy between the curves is another measure of our
uncertainty in $\Delta G(n)$.

\begin{figure}
\hbox to \hsize{\epsfxsize=1.0\hsize\hfil\epsfbox{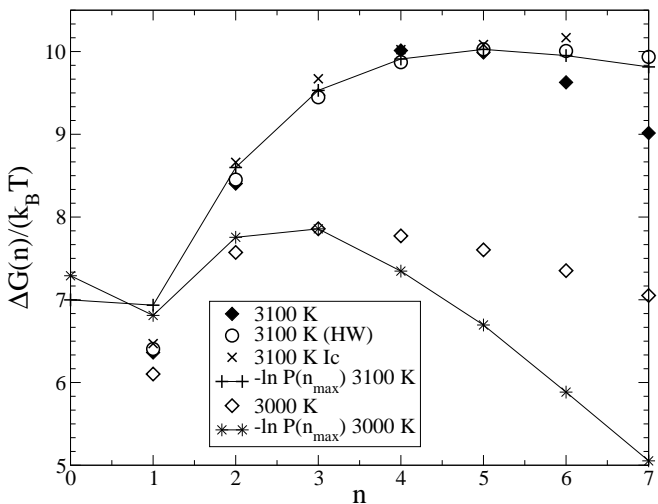}}
\caption{
$\Delta G(n)$ obtained from $N(n)$ and $P(n_{\rm max})$ for
$T=3000$~K, compared to results for $T=3100$~K.  Open and filled
diamonds represent the same data as in Fig.~\ref{fig:profiles}.  
Crosses show results
for $T=3100$~K obtained from simulations in Table~\ref{tab:sim2}(c),
while open circles are from hard wall constraint simulations
described in Table~\ref{tab:sim3}.  All three curves agree up until
the critical size.  The line connecting plus symbols ($+$) shows for
Table~\ref{tab:sim3} the
approximate equality of $P(n_{\rm max})$ and $N(n)$ for $n\ge 2$.
The line connecting stars ($*$) shows that this equality breaks down
for $T=3000$~K.  
}
\label{fig:profiles1}
\end{figure}

As described above, we have calculated $\Delta G(n)$ for $T=3100$~K with
three sets of simulations, as described in Tables~\ref{tab:sim2}(b-c)
and~\ref{tab:sim3}.  In Fig.~\ref{fig:profiles1} we plot as crosses
the results from Table~\ref{tab:sim2}(c), and see that the data show
good consistency with those obtained from Table~\ref{tab:sim2}(b),
deviating only beyond $n=5$, the last $n_0$ in
Table~\ref{tab:sim2}(c).  The hard wall curve also shows good
consistency with the parabolic constraint results.

Fig.~\ref{fig:profiles1} also shows that while $N(n)\approx P(n_{\rm
  max})$ holds for $T=3100$~K, it breaks down for $T=3000$~K,
necessitating the direct calculation of $N(n)$ which we accomplish
with our hard wall constraints.

\begin{figure}
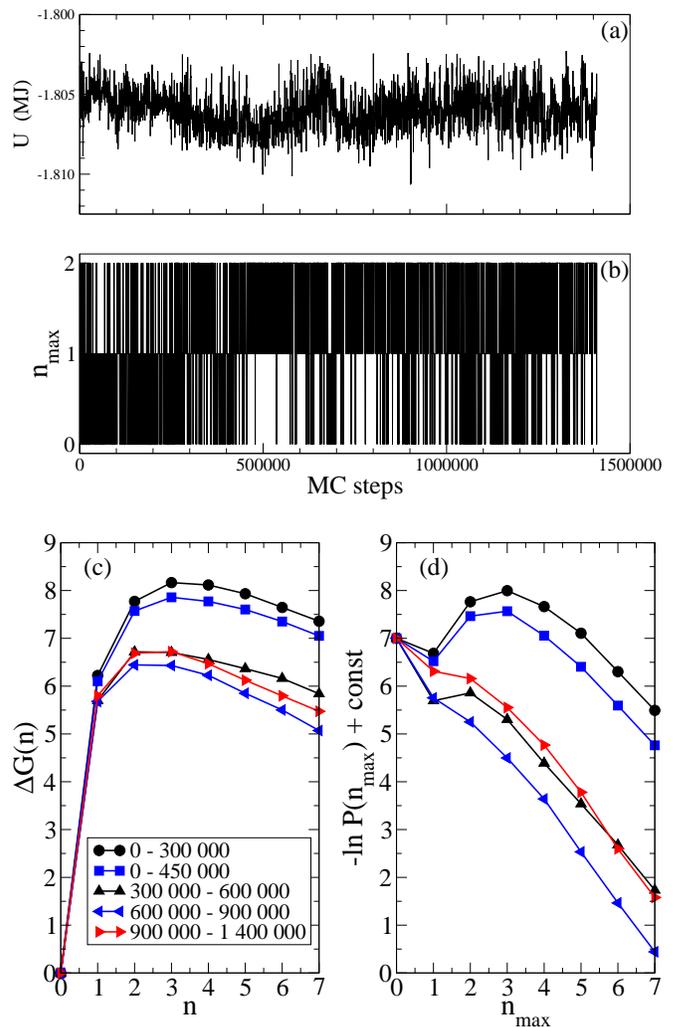

\hbox to \hsize{\epsfxsize=1.0\hsize\hfil\epsfbox{fig9a.eps}}
\vspace{0.4cm}
\hbox to \hsize{\epsfxsize=1.0\hsize\hfil\epsfbox{fig9b.eps}}
\caption{
Breakdown of $n_{\rm max}$ order parameter at $T=3000$~K.  Time series are shown for node $0$ in 
Table~\ref{tab:sim3} for $U$ (a) and $n_{\rm max}$ (b).  In (a), $U$ begins in a steady state similar to the 
unconstrained MD simulations at $3000$~K ($0 - 300\,000$ MC steps), and following period of small decrease ($300\,000-500\,000$ MC steps) enters a regime with slightly larger fluctuations and where lower energy states are probed more often.  Panel (b) shows  a crossover near $400\,000$ MC steps to a regime where $n_{\rm max}=2$ is favored over $n_{\rm max}=0$.  The effect on $\Delta G(n)=-\ln{N(n)}+const$ is shown in (c), where distributions taken from different portions of the time series are plotted: the early time profile is distinct from the later-time steady state.  Panel (d) plots $-\ln{P(n_{\rm max})} + const$, showing it to be monotonically decreasing at late times, apparently indicating a barrierless regime.  However, it is more likely that the nucleation process is simply not adequately described by using $n_{\rm max}$ alone.
}
\label{fig:break}
\end{figure}

However, despite the narrow binning of  $n_{\rm max}$ outlined in Table~\ref{tab:sim3}, all the bins
at $T=3000$~K eventually nucleate.  Fig.~\ref{fig:break} shows for node $0$ in Table~\ref{tab:sim3}
the $U$ and $n_{\rm max}$
time series [panels (a) and (b)] as well as early and late time distributions $N(n)$ [panel (c)]
and $P(n_{\rm max})$ [panel (d)].  The $U$ time series at first glance seems stable, but does
show larger fluctuations at later stages than at the beginning.  The $n_{\rm max}$ time series
also shows a change in behavior: at early times (up to $400\,000$~MC steps) 
$n_{\rm max}= 0$ or $1$ is favored, while at later times $n_{\rm max}= 1$ or $2$ is 
favored.  The early and late time $N(n)$ profiles show a significant difference, as do the
$P(n_{\rm max})$ curves.  In fact, at time beyond $500\, 000$~MC steps, $-\ln P(n_{\rm max})$ 
monotonically decreases.

Since $n_{\rm max}$ is an order parameter, the quantity $-k_{\rm B}T \ln{P(n_{\rm max})}$ is a free energy.  The 
later-time curves shown in panel (d), therefore, would seem to indicate that there is no barrier to increasing $n_{\rm max}$ for the system, i.e. that the liquid is no longer a metastable phase.  
Alternatively, these data might suggest that it is no longer sufficient to describe the nucleation reaction coordinate solely in terms of $n$ and that an additional parameter such as cluster quality may be needed~\cite{moroni}. For example, it could be the case that  small clusters that are well ordered may be over the barrier, as indicated in the second half of the time series, while other less ordered clusters of the same size are on the liquid side of the barrier.
Thus, for late times, this bin no longer contains the metastable liquid, but rather a post-critical state.  
The late-time $\Delta G(n)$ profiles, therefore, represent a
lower bound of the CNT barrier for this $T$.  The early time behavior, however, still shows a metastable liquid
according to $-\ln{P(n_{\rm max})}$, and is therefore used to estimate the CNT barrier.  The curves for 
$T=3100$~K show no such difficulties.

These observations highlight the difficulty in determining $\Delta G(n)$ for such small values of $n^*$, where the quality of the cluster may significantly affect whether a cluster of a given size is post-critical.  A possible solution is to introduce another order parameter that takes into account the quality of the cluster when determining free energy profiles.

Our main result from Fig.~\ref{fig:profiles1} is that we
obtain $\Delta G(n)/k_{\rm B}T=7.86 \pm 0.6$ for $T=3000$~K.  The
uncertainty is obtained by considering different portions of the (early) time
series when constructing $\Delta G(n)$.  We also obtain $n^*=3$ Si
atoms (SiO$_2$ molecules), or $\sim9$ atoms including O.

\subsection{Comparison with CNT}

Fig.~\ref{fig:profiles} shows the full $\Delta G(n)$ curves for the
different $T$.  We see from Fig.~\ref{fig:profiles} that both $n^*$
and $\Delta G(n^\ast)$ decrease as $T$ decreases.  Furthermore, we see
that $n^*$ ranges from 3 to 10.  These small values make it unlikely
that periodic boundary conditions induce catastrophic nucleation in
our system~\cite{honeycutt}.  Moreover, for $T=3000$~K we find $n^*=3$, and
analyzing our MD simulation runs, we confirm that the energy (or
pressure) signature of crystallization occurs after nucleation occurs,
and that the creation of a critical nucleus at this $T$ does not
detectably affect the pressure.

The results in Fig.~\ref{fig:profiles} allow us to compare the
observed behavior to the form of $\Delta G(n)$ predicted by CNT, given
in Eq.~\ref{eq:delgCNT}.  We fit Eq.~\ref{eq:delgCNT} to the data at
each $T$, and determine the constants $a_{\rm fit}$ and $\Delta
\mu_{\rm fit}$ as fitting parameters.  These fits are shown as solid
lines in Fig.~\ref{fig:profiles}, and show that the functional form of
Eq.~\ref{eq:delgCNT} satisfactorily describes the data at all $T$.
For $T=3000$~K, the fit is satisfactory only up to the top portion of
the curve.

\begin{table}
\begin{tabular}{|l|c|c|c|c|}
\hline
 $T$ (K) & $\left| \Delta \mu\right|/k_{\rm B}T$  & $\left| \Delta \mu_{\rm fit} \right|/k_{\rm B}T$ 
   & $a_{\rm fit}/k_{\rm B}T$  & $\gamma/k_{\rm B}T$~{nm}$^{-2}$ \\
\hline
3000              &  3.28  &  5.14  &   11.23  &  29    \\
% 3000 (1-par) & 3.28  & 3.28  &   8.47  &  22 \\
3100  (HW)   &  3.12  &  3.97  &    10.28   &  26    \\
3100              &  3.12  &  4.36  &    10.84  &  24   \\
3200              &  2.96  &  3.57  &     10.00  &  26   \\
3300               &  2.81  &  2.87  &    9.08   &   23   \\
3300 (1-par)  &  2.81  &  2.81  &     8.96   &  23   \\
\hline
\end{tabular}
\caption{Fit parameters using Eq.~\ref{eq:delgCNT} to describe data as plotted in Fig.~\ref{fig:profiles}.  The quantity $\left| \Delta \mu\right|/k_{\rm B}T$ is not a fit parameter, and is determined in a way described in Ref.~\cite{PD}, within an error of $\pm 0.08$.  The  label {\it 1-par} indicates a one-parameter fit in which only $a_{\rm fit}$ is varied.  Estimates of $\gamma$ are obtained from $a_{\rm fit}$,  assuming a spherical nucleus.}
\label{tab:fits}
\end{table}

Next, given that the presence of the critical nucleus does not affect
the system pressure, we calculate $\Delta \mu$ between stishovite and
the liquid at the $T$ and $P$ of the liquid $VT$ state point under
consideration (calculated as the difference in Gibbs free energy per
mole of Si, or SiO$_2$ unit), based on the results presented in
Ref.~\cite{PD}.  The values are presented in Table~\ref{tab:fits},
and compared to the corresponding values of $\Delta \mu_{\rm fit}$.
We find that $\Delta \mu_{\rm fit}$ compares well to $\Delta \mu$ at
$T=3300$~K, but that differences appear at lower $T$, getting larger
as $T$ decreases.  Thus, though the form of Eq.~\ref{eq:delgCNT} fits
the data for $\Delta G(n)$ at all $T$, the ability of CNT to predict
$\Delta \mu$ is lost for $T<3300$~K.

{%\bf
Assuming that we have an approximately spherical nucleus,
we estimate the surface tension from the fit parameter $a_{\rm fit}$
for our range of $T$ to be $\gamma/k_{\rm B}T \approx 25$~{nm}$^{-2}$;
see Table~\ref{tab:fits}.  Note that
$a=4\pi(3/(4\pi\rho_n))^{2/3}\gamma$ for spheres.  
For silica at ambient $P$ and near $T=1500$~K, experimental
values for $\gamma$ range from $0.3$ to $0.7$~Jm$^{-2}$, or
$\gamma/k_{\rm B}T=15$ to $34$~{nm}$^{-2}$~\cite{GAMMA}.  For
comparison, the value recently reported for NaCl at $800$~K is
$\gamma_{\rm NaCl}=80$~erg~cm$^{-2}$, or $\gamma_{\rm NaCl}/k_{\rm
  B}T=7.2$~nm$^{-2}$.  Thus, we see that at our high $P$ and $T$,
where the liquid is simpler, i.e., does not have a tetrahedral
network, $\gamma$ is still close in value to what it is at ambient $P$, 
and does not have a value closer
to that of a simple ionic liquid.  Table~\ref{tab:fits} also shows
that despite the breakdown in the ability of CNT to predict $\Delta
\mu$ in this $T$ range, a fit of Eq.~\ref{eq:delgCNT} to our $\Delta
G(n)$ data still gives a relatively consistent estimate of $\gamma$.

Furthermore, it is interesting to note that while $\Delta \mu_{\rm fit}/k_{\rm B}T$ changes some 80\% as
$T$ decreases from $3300$ to $3000$~K, $\gamma/k_{\rm B}T$ roughly changes by only 
25\%.  This perhaps indicates that the structure and/or density of the critical nucleus interior undergo
larger changes with $T$ than surface properties.

}

\subsection{Kinetic prefactor}

The crucial quantity in the kinetic prefactor is either $\lambda$, or
$f_{\rm crit}^+$ from Eqs.~\ref{eq:K1} and~\ref{eq:K2}.  Following the
work of Frenkel and co-worders~\cite{frenkel_2004,frenkel_2005}, we calculate $f_{\rm
  crit}^+$ through Eq.~\ref{eq:fcrit}.  Eq.~\ref{eq:fcrit} follows the
assumption that the addition and detachment of particles from the
near-critical crystallite is a diffusive process.  In order to measure
the deviation of the cluster size from the critical value, i.e., the
right hand side of the equation, we isolate 80 clusters near the
critical size from constrained MC simulations and use them to seed
$NVE$ simulations lasting $150$~ps with randomized initial velocities
corresponding to $T=3000$~K.  We then use multiple time origins from
each time series, where at each time origin the configuration has
$n_{\rm max}=n^*$.  Additionally, to ensure we are measuring the properties
of clusters of critical size, each time origin is only chosen when the average cluster
size for the preceding $1000$ fs is between $2$ and $4$.  Varying the averaging
time or these upper and lower bounds does not appreciably affect the results.

\begin{figure}
\hbox to \hsize{\epsfxsize=1.0\hsize\hfil\epsfbox{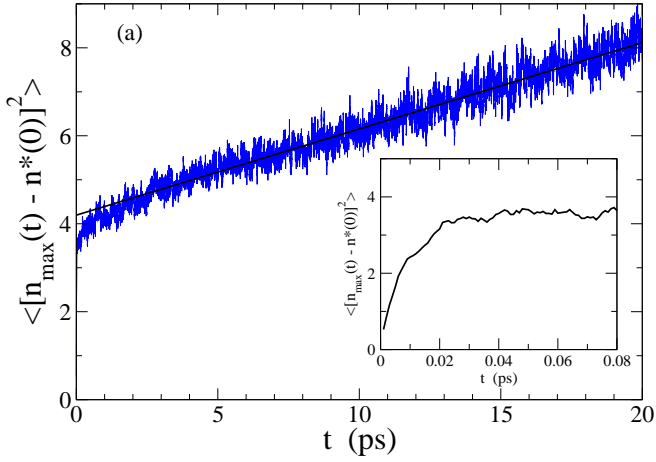}}
\vspace{1.1cm}
\hbox to \hsize{\epsfxsize=1.0\hsize\hfil\epsfbox{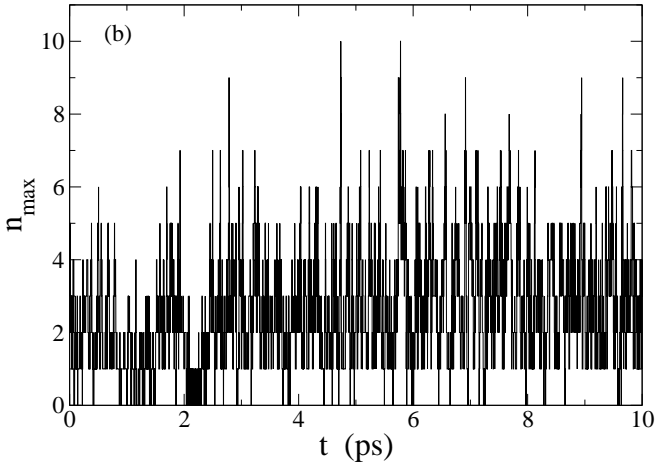}}
\caption{
Calculation of $f_{crit}^+$.
Panel (a) shows a plot of $\left<[n(t)-n^*(0)]^2\right>$ as a function of time at $3000$~K.  After a brief early time regime, the size of the cluster shows diffusive behavior.  The slope of the line of best fit in the linear regime is $(2.0 \pm 0.2) \times 10^2$~{ns}$^{-1} = 2f^+_{\rm crit}$.  Inset shows early time behavior. Panel (b) shows $n_{\rm max}(t)$ for a portion of an $NVE$ simulation seeded witha cluster of size $n^\star=3$.
}
\label{fig:n2}
\end{figure}

We plot in Fig.~\ref{fig:n2}(a) the quantity $\left<\left[n_{\rm
      max}(t)-n^*(0)\right]^2\right>$.  The plot shows a very rapid
early time increase to a value of about $4$ (inset shows early time
behavior) or a fluctuation in size of the cluster of about $2$
particles. 
%Considering that $n^*=3$, this is perhaps alarming and we
%return to this point later.
Notwithstanding the
%magnitude of the
early time change in $\left<\left[n_{\rm max}(t)-n^*(0)\right]^2\right>$,
we see that the time series
enters into a diffusive regime that is linear in time, with
$\left | n_{\rm max}(t)-n^*(0) \right |$ between $2$ and $3$.  By
fitting a line to this section, we obtain an estimate of the slope
$m=(2.0 \pm 0.2) \times 10^2$~{ns}$^{-1}$ which gives $f^+_{\rm
  crit}=m/2=(1.0 \pm 0.1) \times 10^2$~{ns}$^{-1}$.
%, or $0.32$~{ps}$^{-1}$. 
This is about 3 times larger than the value
obtained for molten NaCl at $T=825$~K and atmospheric $P$ of
$0.033$~{ps}$^{-1}$~\cite{frenkel_2005}.

The early time behavior of $\left<\left[n_{\rm max}(t)-n^*(0)\right]^2\right>$
is plotted in the inset of Fig.~\ref{fig:n2}(a), and shows a rapid increase corresponding
to short-time fluctuations in the cluster size.  These rapid fluctuations are seen
in Fig.~\ref{fig:n2}(b), where we plot a representative portion of an $NVE$ simulation
with a critical cluster in it.  Although short-time fluctuations can be considerable given that $n^\star=3$, the general trend shown here suggests $n_{\rm max}$ fluctuates around $n^\star$.

\begin{table}
\begin{tabular}{|c|c|}
\hline
Quantity &  Value \\
\hline
 $n^*$  &   3    \\
 $\Delta G(n^*)/k_BT$ &  $7.86 \pm 0.6$ \\
 $f^+_{crit}$  &  $(1.0 \pm 0.1  )\times 10^2$~{ns}$^{-1}$ \\
 $\left| \Delta \mu \right|/k_BT$  &  $3.28 \pm 0.08$ \\
 $\rho_{n}$  & $43.8929$~{nm}$^{-3}$ \\
 $D$ & $(8.0 \pm 0.2)\times 10^{-8}$~{nm}$^2${fs}$^{-1}$ \\
 $Z$  &   0.241 \\
 $ \lambda $ &  0.2~nm \\
 \hline
 \hline
 $ J $  &  $6\times10^{34}$~m$^{-3}$s$^{-1}$ \\
 $ J^{CNT}$  &  $4.1\times 10^{35}$~m$^{-3}$s$^{-1}$ \\
 $ J^{ms}$  &  $1.6\times 10^{34}$~m$^{-3}$s$^{-1}$ \\
\hline
\end{tabular}
\caption{Summary of calculated quantities for $T=3000$~K.}
\label{tab:q}
\end{table}

All the factors required to calculate the nucleation rate via
Eq.~\ref{eq:JCNT} are summarized in Table~\ref{tab:q}.  The resulting
rate is $J^{CNT}=4.1\times 10^{35}$~m$^{-3}$s$^{-1}$, and given the
uncertainties in the calculated quantities, this result should be
accurate within a factor of 2.  Note that we have calculated the rate
using $\left|\Delta \mu \right|$ as obtained from independent free
energy calculations. It could be argued that $\left|\Delta \mu_{\rm
    fit} \right|$ is the appropriate quantity and this introduces an
additional factor of uncertainty of $\sqrt{\left|\Delta \mu_{\rm fit}
  \right|/\left|\Delta \mu \right|}=1.25$

%and not $\left|\Delta \mu_{\rm fit} \right|$ as obtained from fits of the profiles.  This ambiguity of which one use introduces a further uncertainty of only a factor of $\sqrt{\left|\Delta \mu_{\rm fit} \right|/\left|\Delta \mu \right|}=1.25$

The quantity $\lambda$ can be obtained by solving Eqs.~\ref{eq:K1}
and~\ref{eq:K2}, resulting in,
\begin{equation}
\lambda=\sqrt{\frac{24D{n^*}^{2/3}}{f^+_{\rm crit}}}.
\end{equation}
Using our values for $D$, $f^+_{\rm crit}$ and $n^*$, we obtain $\lambda=0.20$~nm. To put this in perspective,  the first peak of the Si-Si radial distribution function for the liquid at our state point is $\approx 0.3$~nm
% This makes $\lambda$ comparable to what one might expect from the
% Lindemann criterion.
 and the width of the first neighbor peak in the liquid $g_{\rm SiSi}(r)$ is about $0.1$~{nm}.

\subsection{Metastable equilibrium liquid $N(n)$}

We now focus on the steady-state liquid that exists in our MD runs
prior to any crystallization, which we call the metastable equilibrium liquid.
We ask whether we can extract information about the barrier to
nucleation from this metastable equilibrium liquid, i.e., without any constrained MC
sampling~\cite{mousseau}.  
To this end, we harvest configurations from our direct
nucleation MD runs that have $500~{\rm ps} \le t \le t_x-500~{\rm
  ps}$.  This reasonably ensures that we have configurations only from
metastable liquid equilibrium, or steady state liquid, and have not
included configurations that have begun to crystallize.  We measure
the distributions of cluster sizes to obtain $N^{ms}(n)$
%This is not the ``steady state nucleation'' distribution referred to in CNT, but rather the distribution from the (metastable) equilibrium liquid.  From $N^{ms}(n)$, we 
and then define the free energy
\begin{equation}
\Delta G^{ms}(n)=-k_{\rm B}T\ln N^{ms}(n) + const,
\end{equation}
where the constant is chosen to ensure $\Delta G^{ms}(0)=0$.  
{%\bf 
One might expect the two distributions $N(n)$ and $N^{ms}(n)$ to be the same.
However, $N(n)$ is obtained by allowing otherwise unstable clusters to equilibrate
in the surrounding liquid, while $N^{ms}(n)$ is obtained directly from a dynamic
simulation.  Furthermore, the way we obtain $N^{ms}(n)$ reduces sampling of
states near the top of the barrier.
}

\begin{figure}
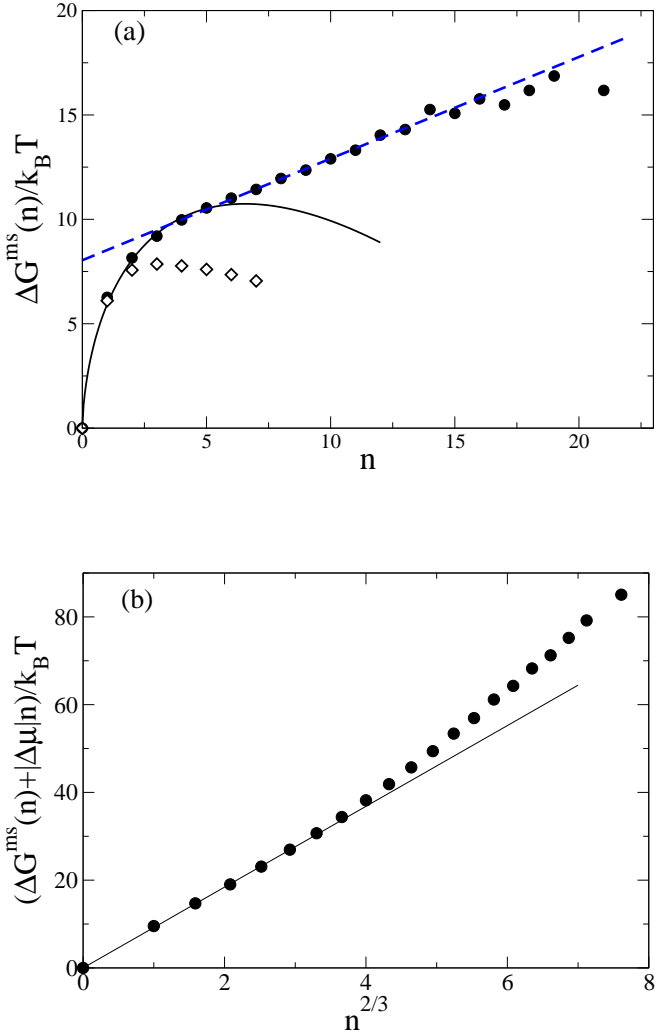

\hbox to \hsize{\epsfxsize=1.0\hsize\hfil\epsfbox{fig11a.eps}}
\vspace{1.3cm}
\hbox to \hsize{\epsfxsize=1.0\hsize\hfil\epsfbox{fig11b.eps}}
\caption{
Work of cluster formation at $T=3000$~K derived from $N^{ms}(n)$ distributions obtained directly from $NVT$ MD simulations where no crystal nucleation has occurred.  
Panel (a) shows $\Delta G^{\rm ms}(n)/k_{\rm B}T$ (filled cirles) along with the equilibrium $\Delta G(n)/k_{\rm B}T$ (open diamonds), the curve $(-\left| \Delta \mu \right| n + 9.2 n^{2/3})/k_{\rm B}T$ (solid curve),
and a fit of the linear part of  $\Delta G^{\rm ms}(n)/k_{\rm B}T$ at high $n$.
Panel (b)  shows the quantity $[\Delta G^{\rm ms}(n) + \left| \Delta \mu \right|n]/k_{\rm B}T$, that according to Eq.~\ref{eq:delgCNT} should be linear function of $n^{2/3}$.  The line of best fit passing through the origin and through the data points corresponding to $n=1$, 2, 3 and 4, has slope 9.2.
}
\label{fig:ss}
\end{figure}

We plot
$\Delta G^{ms}(n)$ in Fig.~\ref{fig:ss}(a) (filled circles).  Although
the first part of the data behaves like a usual barrier, the curve
at larger $n$ is straight.  A linear $\Delta G^{ms}(n)$ implies an
exponential $N^{ms}(n)$.  It also implies that the work required to
add a particle to the cluster is independent of $n$.  Clearly, this
does not fit the CNT picture, where increasing the size of the cluster
reduces the work required to add a particle.

In order to extract some information from $\Delta G^{ms}(n)$, we
calculate the quantity $[\Delta G^{ms}(n)+\left|\Delta \mu\right|
n]/k_{\rm B}T$, which should be a linear function of $n^{2/3}$, and
plot it in Fig.~\ref{fig:ss}(b).  The resulting curve does indeed show
a linear dependence on $n^{2/3}$ at small $n$.  A fit through the
origin and the first 4 non-zero data points yields a slope of $9.2$.
Using this slope, we construct the CNT curve $C^{ms}(n)=-\left| \Delta
  \mu \right| n + 9.2 n^{2/3}$ and plot it in Fig.~\ref{fig:ss}(a)
(solid line).  From this curve, we obtain $\Delta G^{ms}(n^*)=10.74$
and $n^*\approx 6$.  Using these parameters, while keeping 
%the dynamic ones 
$f^+_{\rm crit}$ calculated earlier, we obtain an estimate of the nucleation rate
of $J^{ms}=1.6\times 10^{34}$~s$^{-1}$m$^{-3}$,
a value closer to $J$ than that obtained from $N(n)$ 
(see Table~\ref{tab:q}).

This close agreement with $J$ still leaves us with question
of why $\Delta G^{ms}(n)\ne \Delta G(n)$.
%, even for small $n$.  
In Fig.~\ref{fig:ss}(a), we plot for comparison $\Delta G(n)$, and see
that is it significantly lower than $\Delta G^{ms}(n)$.

\section{Discussion}

In this study, we take advantage of a liquid state point that spontaneously 
nucleates on a time scale long enough to allow the determination of the properties 
of the metastable liquid.  By sampling many nucleation events, we obtain the rate 
directly as shown in Fig.~\ref{fig:rate}, where we show that the nucleation process
enters a regime of first-order kinetics.  The criterion used to determine when nucleation
has taken place, whether an energy criteria or an examination of the cluster size, does
not significantly alter our estimate of the rate.

With the direct rate in hand, we wish to test the prediction of CNT.  First, we calculate
$\Delta G(n)$ for a series of $T$ above and including $T_0$.  
As it is defined in Eq.~\ref{eq:delg}, $\Delta G(n)$ is only formally 
a (relative) free energy for the case when $n$-sized clusters are rare.  In such a case,
$\Delta G(n)$ becomes equivalent to 
$-\ln P(n_{\rm max})$, which is formally a free energy.
However, $N(n)$ is the quantity of central
importance in CNT, and the interpretation of $\Delta G(n)$ as a free energy,
and $\Delta G(n^*)$ as a free energy barrier, is valid for more moderate 
supercooling or for liquid condensation from the supersaturated vapor.

In this work, we probe $T$ low enough that a considerable (small-$n$) portion of 
$N(n)$ is not equivalent to $P(n_{\rm max})$.  Indeed, at $T=3000$~K, where the small
value of $n^*$ makes all cluster sizes of interest sufficiently common, $N(n)$ and 
$P(n_{\rm max})$ are different everywhere.  This does not affect the formalism of CNT,
merely the identification of $\Delta G(n)$, as defined in Eq.~\ref{eq:delg}, as a free energy.

This becomes important at $T=3000$~K, where it is difficult to keep the metastable liquid
from nucleating, even at the smallest range of $n_{\rm max}$.  At this $T$, the free energy
barrier according to $P(n_{\rm max})$ is about $1.5$ $k_{\rm B}T$ [Fig.~\ref{fig:break}(d)].
The possibility of a spinodal-like loss of liquid stability to the crystal becomes a prominent
possibility~\cite{spinodal}.  On the other hand,
a set of criteria are used in order to distinguish liquid and crystal states.
Thus, in a system where nucleation is taking place in a localized region, most of the system
will still be labeled as liquid.  Therefore, there will be more liquid-like
particles (corresponding to $n=0$) than single crystal-like particles (corresponding to $n=1$),
so that $\Delta G(n)$ will always show a barrier.  Hence, if nucleation is unavoidable
because of some spinodal process, the usefulness of interpreting $\Delta G(n^*)$
as a free energy barrier is not clear.

With these thoughts in mind, we proceed to discuss the progression of $\Delta G(n)$ with $T$.
At $T=3300$~K, the identification of $N(n)$ with $P(n_{\rm max})$ holds very well.  The resulting
$\Delta G(n)$ is well described by Eq.~\ref{eq:delgCNT}.  Indeed, even the bulk value for
$\Delta \mu$ calculated independently accurately describes the data, necessitating a fit only
to determine the surface tension term.  It is important to note that our approximation that the
appearance of the critical nucleus does not affect P should be weakest at this highest
$T$ since the critical nucleus is largest.  The $\Delta \mu$ result upholds our approximation.

The $\Delta G(n)$ profiles for $T=3200$ and $3100$~K are similar to the $3300$~K case, 
except that the bulk value of $\Delta \mu$ shows increasing deviation from 
$\Delta \mu_{\rm fit}$.  This deviation stems from either violating the assumption of
cluster incompressibility used in the derivation of Eq.~\ref{eq:delgCNT}~\cite{pablo,reiss}, 
or from the
(near certain) possibility that the structure of the small nuclei is different from bulk
stishovite, and therefore follows a different equation of state (the bulk equation of state is
used calculate $\Delta \mu$).

Here, a note about ensembles is in order.  Since we are at constant $V$,
$\Delta G(n)$ [or $-\ln P(n_{\rm max})$ for that matter] represent a change in 
Helmholtz free energy of the system.  The use of $\Delta \mu$ in Eq.~\ref{eq:delgCNT},
however, is still formally correct~\cite{bowles_2003}.  Moreover, as we have shown, the critical 
nucleus does not significantly alter the $P$ in our system.  Hence, the nucleus can
be regarded as being in either a constant $V$ or a constant $P$ environment.

At $T=3000$~K, clusters are insufficiently rare to identify $N(n)$ with $P(n_{\rm max})$.
Therefore, we calculate $N(n)$ directly using hard-wall constraints, a method giving
the same results for $3100$~K as those obtained from the parabolic constraint.
Furthermore, we see a breakdown in the ability of our single order parameter
$n_{\rm max}$ to sufficiently characterize the critical cluster.  A high quality cluster
of size $2$ can be post-critical, resulting in what appears to be a spinodal-like profile
in Fig.~\ref{fig:break}(d) for late times.  For our current purposes, we use the portion 
of the time series which explicitly shows the metastable liquid state in order to calculate
$\Delta G(n)$.  In order to calculate $P(n_{\rm max})$ accurately at $T$ near $3000$~K, more
stringent measures should be taken, including perhaps using an extra order parameter to help
characterize the critical state better.

Calculating the kinetic components of the CNT expression for the rate, though more
straightforward, requires comment.
The parameter $\lambda$ is usually defined as the distance a particle must diffuse when moving from the surrounding fluid to the nucleating phase, which is an intuitive interpretation in the case of condensation in a dilute gas. In the case of crystal nucleation, its meaning is not so clear, especially when we ask how $\lambda$ should be interpreted with respect to a cluster criteria which identifies  correlations between particle environments. Nevertheless, by calculating $D$ and $f^+_{\rm crit}$, we find $\lambda=0.2$~nm. This value is physically appealing, as it is less than the first neighbor Si-Si distance of $0.3$~nm and the distance between the two sub peaks of the first $g_{\rm SiSi}(r)$ peak for stishovite (the first peak is split), is about $0.1$~nm.  
This reasonable value of $\lambda$ is evidence that CNT provides an adequate description of nucleation
in our system, a contrast to the case of  molten NaCl, where $\lambda$ was found to be unphysically 
large~\cite{frenkel_2005}. 

Our calculation of $f^+_{\rm crit}$ suggests there are two time scales associated with the dynamics of the critical cluster. $[n_{\rm max}(t)-n^*(0)]^2$ grows rapidly over the first 40~fs  before reaching a plateau near 4, after which, it increases at a much slower rate. The diffusive growth of the cluster occurs slowly and measurements of $f^+_{\rm crit}$ on the longer time scale leads to reasonable values of $\lambda$. The short-time fluctuations most likely arise arise from particles close to the cluster definition thresholds, for which small motions result in their being included or excluded in the crystalline cluster.

We emphasize that despite the difficulties at $3000$~K, the rates $J$ and $J^{CNT}$ compare quite
favorably, given similar comparisons done previously~\cite{frenkel_2005}.  At higher $T$, there 
are no difficulties with the formalism used to calculate $\Delta G(n)$.  Furthermore, the quantitative
agreement between $\Delta G(n)$ and Eq.~\ref{eq:delgCNT} is excellent, especially given the fact 
that $\Delta \mu$ is calculated independently for bulk phases.

Another intriguing aspect of this study is the difference between $\Delta G(n)$ and $\Delta G^{ms}(n)$.
$\Delta G(n)$ is obtained through equilibrium simulations of the constrained system.  The constraint allows for a rigorous determination on $N(n)$, allowing  post-critical states to be sampled while determining the barrier.  $\Delta G^{ms}(n)$ is determined through MD simulations of liquid quenched from $T=5000$~K to $T_0$, allowing the liquid to relax for several $\alpha$-relaxation times, and collecting data only until $500$~ps before crystallization is detected through the energy.  
Only three runs have $t_x-t_{\rm nuc} > 500$~ps, with the largest difference being $781$~ps.  Increasing the cutoff of the time series to $800$~ps before $t_x$ does not significantly alter $\Delta G^{ms}(n)$.

Thus, $\Delta G^{ms}(n)$ is constrained in a peculiar way.  The data are pruned to include post-critical structures, so long as they happen to dissolve through some fluctuation.  Therefore, post-critical states are sampled in a non-equilibrium fashion.  Also,  pre-critical fluctuations that happen to carry the system over the barrier quickly are not sampled well either.  Therefore, near-critical states are sampled less often than in equilibrium [and hence $\Delta G^{ms}(n) > \Delta G(n)$].  We have not determined whether this sampling difference is sufficient to account for the difference between $\Delta G^{ms}(n)$ and $\Delta G(n)$.

Given these difficulties, the agreement between $J$ and the rate calculated from $N^{ms}(n)$, following from our seemingly logical procedure to obtain a barrier height from $\Delta G^{ms}(n)$ [$C^{ms}(n^*)$],
may be fortuitous.  However, another possibility regarding the discrepancy between $\Delta G^{ms}(n)$ and  $\Delta G(n)$ is that the time  scale on which the $N(n)$ evolves is much longer that the $\alpha$-relaxation time of the liquid.  We note that quantities like $g(r)$ and the structure factor are constant during the time $500$~ps $ < t < t_x - 500$~ps.  However, it is possible that $N(n)$ evolves more slowly.  %There is evidence suggesting this: systems quenched from different initial equilibrium $T$ to the same undercooling can experience different rates of nucleation~\cite{KKKK}.
In this case, the $N^{ms}(n)$ that we measure is not expected to be the same as $N(n)$, and is perhaps more physically relevant in calculating the rate.  Perhaps this is why $J^{ms}$ agrees better with $J$ than does $J^{CNT}$.  Clearly, the comparison of the metastable liquid distribution and the equilibrium distribution raises a number of interesting questions that warrant further investigation.

%We also do not know what  effect our thermostat has on $N^{ms}(n)$.  Previous work has shown that the thermostat used does not seriously affect the cluster size distribution~\cite{tanaka2005}. IS THIS IMPORTANT?

\section{Conclusions}

We perform $NVT$ MD simulations of liquid silica at $T_0=3000$~K and $V_0=4.5733$~cm$^3$/mol, corresponding to $P_0=44.0$~GPa, and calculate the rate of homogenous nucleation to stishovite to be $J=(6.0 \pm 0.2) \times 10^{34}$~{m}$^{-3}${s}$^{-1}$.  This state point is located deep in the stishovite field in the $PT$ phase diagram, and within the single phase coexistence region of stishovite in the $VT$ phase diagram.  $T_0$  is at about half the melting temperature at this $P_0$.  

We also compare this rate to that predicted by CNT.  The work in forming a cluster of size $n$ [$\Delta G(n)$] follows the form predicted by CNT (Eq.~\ref{eq:delgCNT}) for $T=3000$~K, $3100$~K, $3200$~K and $3300$~K. At $3300$~K, an independent calculation of $\Delta \mu$ using bulk phase values allows for a successful one-parameter fit of  $\Delta G(n)$.
Assuming a spherical nucleus, an estimate for surface tension in this range of $T$ is $\gamma/k_{\rm B}T \approx 25$.
At $T=3000$~K, the CNT form only fits the data well only up to $n=4$, one larger than the critical size.  Furthermore, the usual identification $N(n)\approx P(n_{\rm max})$ breaks down strongly at this $T$. 

We also find that $N(n)$ and $N^{ms}(n)$ differ.  In fact, using $N^{ms}$ yields a CNT result that is closer to direct measurement.  This may indicate a subtle dependence of the nucleation rate on the initial $T$ from which the system is quenched, via slow evolution of the cluster distribution.

Calculating the kinetic prefactor, we obtain $f^+_{\rm crit}= 100 \pm 10$~{ns}$^{-1}$, resulting in a calculated rate of $4.1 \times 10^{35}$~m$^{-3}$s$^{-1}$, with an uncertainty of a factor of $2$.  Therefore CNT overestimates the rate by an order of magnitude.  The average distance that a Si atom must diffuse in attaching itself to a crystalline cluster is $\lambda=0.2$~nm.  This length is approximately twice the width of the first neighbor shell of the Si-Si radial distribution function.%, or the distance between the split peaks of the fisrt neighbor SiSi shell of stishovite.This value of $\lambda$ is physically quite reasonable.

%Thus, we conclude by saying that CNT adequately describes nucleation in our system of study, both qualitatively and quantitatively.

\section{Acknowledgments}
We thank the StFX hpcLAB and G.~Lukeman for computing resources and support,
and acknowledge support from the AIF, the CFI, the CRC Program, and NSERC.

\section{Appendix: Determining cluster size}

\begin{table}
\begin{tabular}{|l|r|c|c|c|}
\hline
Structure & $N_b$  & $Q_4$ & $Q_6$  &  $Q_8$ \\
\hline
FCC            &  12  &  0.19  &  0.57  &  0.40  \\
BCC            &  12  &  0.08  &  0.54  &  0.38  \\
HCP            &  12  &  0.10  &  0.48  &  0.32  \\
SC               &    6  &  0.76  &  0.35  &  0.72  \\
SC               &  10  &  0.40  &  0.02  &  0.60  \\
LIQ              &   10 &  0.02  &  0.03  &  0.02  \\
X-LIQ          &    6  &  0.21  &   0.33  &  0.24  \\
X-LIQ          &    8  &  0.23  &   0.30  &  0.27  \\
X-LIQ          &  10  &  0.23  &   0.27  &  0.29  \\
X-LIQ          &  12  &  0.11  &  0.22  &  0.33  \\
ST 3000K   &    6  &  0.39  &  0.52  &  0.33  \\
ST 3000K   &    8  &  0.39  &  0.48  &  0.35  \\
ST 3000K   &  10  &  0.40  &  0.45  &  0.38  \\
ST 3000K   &  12  &  0.25  &  0.38  &  0.42  \\
ST  0K         &   10 &  0.41  &  0.51  &  0.42  \\
\hline
\end{tabular}
\caption{$Q_l$ for $l=4$, 6 and 8, for various structures and choices of $N_b$: face-centered cubic (FCC), body-centered cubic (BCC), hexagonally close-packed (HCP), simple  cubic (SC), liquid silica at $T=3000$~K (LIQ), stishovite (ST) at $3000$~K, stishovite at $0$~K, and the structure that results when the liquid spontaneously crystallizes to stishovite at $3000$~K (X-LIQ).}
\label{tab:qvalues}
\end{table}

\begin{figure}
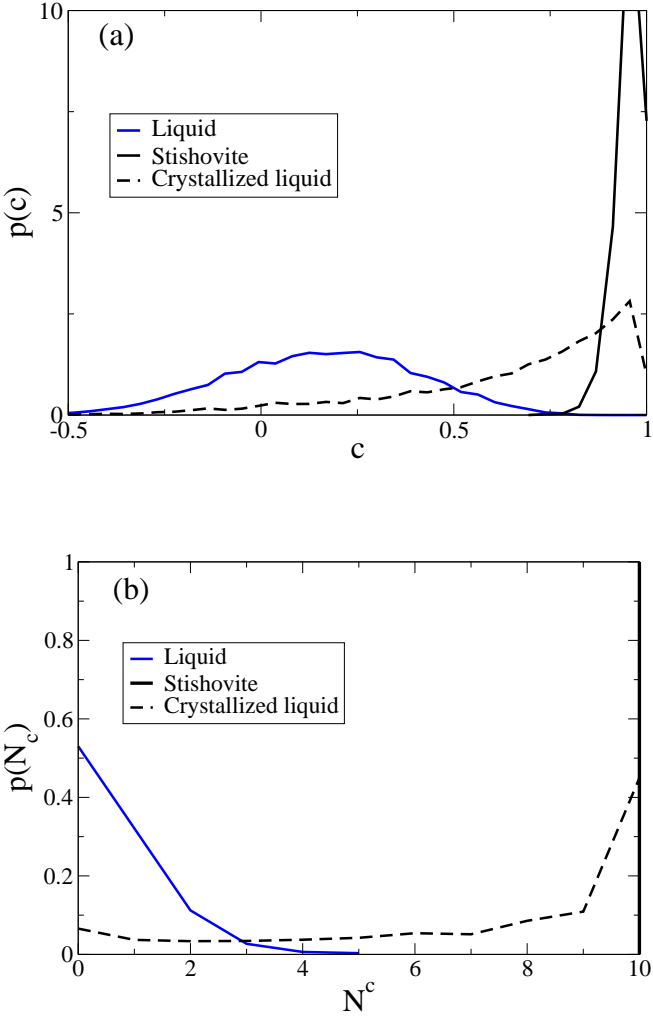

\hbox to \hsize{\epsfxsize=1.0\hsize\hfil\epsfbox{fig12a.eps}}
\vspace{1.2cm}
\hbox to \hsize{\epsfxsize=1.0\hsize\hfil\epsfbox{fig12b.eps}}
\caption{
Determination of characterization thresholds.  We plot the probability distribution of $c_{ij}$ values in (a) for the liquid, stishovite and the spontaneously crystallized liquid at $T_0$ and $V_0$.  The results in each case are averages from five configurations.  Based on this plot, we choose a value of $c_{\rm cut}=0.5$.  In (b) we plot the probability distribution of $N^c$ values based on $c_{\rm cut}$.  At this $T$, stishovite appears to always have ten Si-Si connections per Si.  Based on this plot, we choose a value of $N^c_{\rm cut}=5$, at or above which a Si ion is deemed to be crystal-like.  
}
\label{fig:qdot}
\end{figure}

We only consider Si atoms in our sample when determining crystallinity.  There are three reasons for this: Si and O atoms have different local geometry and so the analysis is made easier by looking only at Si-Si structure; it is computationally faster to do so; and our order parameter does not track, and therefore does not influence, what the O atoms are doing, allowing for greater structural freedom during the constrained MC simulations.

In terms of determining local geometry,
the difficulty arises in describing the different environments of Si and O
atoms.  In the case of NaCl, both species have the same coordination
environment in the solid, and therefore can be described with one scheme.
In the case of SiO$_2$, rather than finding a way of describing Si-O, O-O and Si-Si
bonds separately, we choose to account for Si atoms only.  This is also
computationally faster.  We assume that the strong local stoichiometry will
persist in the growing clusters as well.

In order to define a crystalline cluster forming within the liquid, we follow the procedure laid out in~\cite{frenkel_1996}.  We need a bond order parameter that captures crystal structure.  To begin with, we use spherical harmonics $Y_{lm}(\hat{r}_{ij})$, where $\hat{r}_{ij}$ is a unit vector pointing along a bond between particles $i$ and $j$ (and thus providing elevation and azimuth angles with respect to a fixed coordinate system).  For FCC and BCC crystals, $l=6$ has been used, while for salt (having cubic structure), $l=4$ has been used~\cite{frenkel_2005}.  The first step is to define a local quantity on the particle level,
\begin{equation}\label{eq:qlm}
%q_{lm}(i)=\frac{1}{N_b(i)}\sum_{j=1}^{N_b(i)}Y(\hat{r}_{ij})
q_{lm}(i)= \sum_{j=1}^{N_b(i)}Y_{lm}(\hat{r}_{ij}),
\end{equation}
where the sum is over the $N_b$ bonds of particle $i$.
A global measure of the overall crystallinity can be written,
\begin{equation}
Q_{lm} =  \frac{\sum_{i=1}^{N}q_{lm}(i)}{\sum_{i=1}^N N_b(i)},
\end{equation}
and hence a quantity that does not depend on the coordinate system is,
\begin{equation}
Q_l = \left( \frac{4\pi}{2l+1} \sum_{m=-l}^l \left| Q_{lm}\right|^2 \right)^{1/2}.
\end{equation}

Usually, $N_b$ is determined via a cutoff distance near the first minimum in the radial distribution function.  Ideally, FCC, BCC and HCP structures have twelve neighbors, while simple cubic has six.  However, during a simulation, $N_b$ will fluctuate.  In the present work, instead of defining a distance cut-off, we always choose the closest ten silicon neighbors of a given silicon atom, i.e., $N_b=10$ always.
In stishovite, the first Si-Si neighbor shell contains ten atoms, although the shell is split with two neighbors slightly closer than the other eight.

In Table~\ref{tab:qvalues}, we list the $Q_l$ values for $l=4$, 6 and 8, for various crystal structures as well as for both the metastable liquid and the state after the liquid has spontaneously crystallized (crystal with defects).  We see that both $Q_6$ and $Q_8$ give high values for most crystals.  However, $l=8$ seem to be less sensitive to the value of $N_b$ chosen.  In particular, for $N_b=10$ in the case of the simple cubic structure, $Q_6$ fares much worse that $Q_8$.  For our study, we do not know what the structure of precritical nuclei of stishovite is, and thus we prefer to have an order parameter that is more accepting of different structures.  Therefore, we choose $l=8$

Having selected $N_b=10$, and $l=8$, we now proceed to determine what a crystal-like atom is, and whether two crystal-like atoms are part of the same cluster.  Having defined $q_{lm}$ in Eq.~\ref{eq:qlm}, we can form a dot product $c$ ($-1\le c\le 1$) between two neighboring Si atoms $i$ and $j$,
\begin{equation}
c_{ij} = \sum_{m=-8}^8 \hat{q}_{8m}(i) \hat{q}^*_{8m}(j),
\end{equation}
where
\begin{equation}
\hat{q}_{8m}(i) = \frac{q_{8m}(i)}{\left(\sum_{m=-8}^8 \left| q_{8m}(i)\right|^2  \right)^{1/2}},
\end{equation}
and $q^*$ is the complex conjugate of $q$.  In this way, $c$ is determined for every pair of neighboring atoms.  For two atoms with very similarly oriented bonding geometry, $c$ will have a value close to unity.

The distribution of $c$ values is plotted in Fig.~\ref{fig:qdot}(a) for stishovite, liquid and spontaneously crystallized liquid all at $T_0$ and $V_0$.  We see from the plot that very few atoms pairs in the liquid have a value greater than about $0.75$, while very few atom pair in stishovite have a value less than $0.75$.  However, a value of $c=0.5$ provides a better criterion for differentiating between the liquid, and the spontaneously crystallized configuration.  Therefore, we choose a cut-off value of $c_{\rm cut}=0.5$.  A pair of neighboring atoms $i$ and $j$ that have $c_{ij}\ge c_{\rm cut}$ are considered to be connected by a crystal-like bond.

To define a crystal-like atom, we say that the number of connections $N^c$ an atom possesses must be greater than or equal to $N^c_{\rm cut}$.  To determine $N^c_{\rm cut}$, we plot the distribution of $N^c$ in Fig.~\ref{fig:qdot}(b) for the same cases as for $c$.  We see that all atoms in the stishovite crystal have $N^c=10$, while the distribution for the  liquid vanishes near $N^c=5$.  From the plot, any value between $5$ and $10$ would serve to distinguish the liquid from the crystal.  We choose $N^c_{\rm cut}=5$ to be less restrictive in our choice of order parameter.  Beyond this, clusters are defined by considering connections only between crystal-like atoms.

\end{document}